\documentclass[a4paper,11pt]{article}
\usepackage{jheppub}

\usepackage{subcaption}

\usepackage{amssymb}
\usepackage{hyperref}
\usepackage{amsmath}
\usepackage{xcolor}
\usepackage{setspace}
\usepackage{enumitem}
\usepackage{graphicx}
\usepackage{tikz}

\usepackage[T1]{fontenc} 
\usepackage{mathtools}
\usepackage{slashed}
\usepackage[section]{placeins}
\usepackage{braket}
\usepackage{upgreek}
\usepackage[bottom]{footmisc}

\newcommand{\eq}[1]{eq.~\eqref{#1}}
\newcommand{\eqs}[2]{eqs.~\eqref{#1} and \eqref{#2}}

\newcommand{\df}{\mathrm{d}}

\newcommand{\nn}{\nonumber}

\DeclareUnicodeCharacter{2032}{'}

\begin{document}

\title{Higher-point Energy Correlators: Factorization in the Back-to-Back Limit \& Non-perturbative Effects}

\author[a]{Ankita Budhraja,}

\author[a,b]{Isabelle Pels,}

\author[a,b]{Wouter J.~Waalewijn}

\affiliation[a]{
Nikhef, Theory Group, Science Park 105, 1098 XG, Amsterdam, The Netherlands
}
\affiliation[b]{
Institute for Theoretical Physics Amsterdam and Delta Institute for Theoretical Physics,
\\
University of Amsterdam, Science Park 904, 1098 XH Amsterdam, The Netherlands
}

\emailAdd{abudhraj@nikhef.nl}
\emailAdd{isabellepels01@gmail.com}
\emailAdd{w.j.waalewijn@uva.nl}


\abstract{$N$-point energy correlators are powerful observables for studying strong interactions, with applications ranging from extractions of the strong coupling $\alpha_s$ to probes of jet modification in heavy-ion collisions and determination of the top-quark mass.
Their practical use has, however, been limited by the complicated phase space for large $N$. Using a recently introduced parametrization that simplifies this structure, we study projected $N$-point correlators in two regimes: factorization in the back-to-back limit and leading non-perturbative effects in the collinear limit. While results in the back-to-back regime were previously limited to the energy-energy correlator, our approach allows us to derive the factorization theorem for arbitrary $N$. We compute the new ingredient, a one-loop jet function, needed for the next-to-next-to-leading-logarithmic resummation, which enables future $\alpha_s$ extractions with complementary systematics.
We further determine the analytic structure of leading non-perturbative power corrections for arbitrary $N$, including their dependence on the center-of-mass energy $Q$, the value of $N$, and the angular scale $x$. We present the first results for non-integer $N<1$, finding that the classical scaling in $x$ acquires an $N$-dependent modification, and that a new non-perturbative matrix element $\tilde\Omega^{[N]}$ appears. In a certain approximation, $\tilde\Omega^{[N]}$ can be related to the standard parameter $\Omega_1$ relevant for $N>1$. Our analytic predictions are tested against the hadronization model in {\textsc{Pythia}}, finding good agreement. The results presented in this paper demonstrate the significant advancements enabled through our new parametrization of energy correlators.
\begin{description}
\item[Keywords:]
Energy Correlators, Jet Substructure, Quantum Chromodynamics
\end{description}
}

\maketitle

\section{Introduction}

Energy correlators have emerged as powerful tools for understanding the underlying energy flow in particle collision experiments, providing unique insights into both the perturbative and non-perturbative aspects of strong interactions~\cite{Moult:2025nhu, Komiske:2022enw, Electron-PositronAlliance:2025fhk}. Originally introduced in the 1970s as clean tests of perturbative quantum chromodynamics (QCD)~\cite{Basham:1977iq,Basham:1978bw, Basham:1978zq, Basham:1979gh}, these observables have experienced a remarkable renaissance in recent years, driven both by the exceptional performance of modern collider detectors and significant theoretical advances in understanding their structure~\cite{Hofman:2008ar,Belitsky:2013xxa,Henn:2019gkr,Dixon:2019uzg,Chen:2019bpb,Luo:2019nig,Chen:2020vvp,Chen:2021gdk}. Energy correlator observables, especially the simplest two-point energy-energy correlator (EEC), have now been measured across a wide range of collision systems, from electron-positron ($e^+\, e^-$) annihilation~\cite{Bossi:2024qeu,Electron-PositronAlliance:2025wzh} to proton-proton~\cite{CMS:2024mlf,ALICE:2024dfl,ALICE:2025igw,STAR:2025jut} and heavy-ion collisions~\cite{CMS:2025ydi,CMS:2024ovv,CMS:2025jam}. The ratio of the 3- and 2-point correlator has the potential to provide the most accurate $\alpha_s$ extraction from jet substructure measurements~\cite{CMS:2024mlf}. Precise computation of energy correlators on charged particles has also been achieved using the track function formalism, thereby enabling significant gains from the angular resolution of the tracking detectors~\cite{Chang:2013rca,Chang:2013iba,Li:2021zcf,Jaarsma:2023ell,Jaarsma:2025tck, Electron-PositronAlliance:2025fhk}. Beyond these, energy correlators have found numerous phenomenological applications, including the determination of the top quark mass~\cite{Holguin:2022epo, Holguin:2023bjf, Holguin:2024tkz, Xiao:2024rol, Holguin:2026vld, Gao:2026xuq}, imaging the hadronization transition~\cite{Komiske:2022enw, Chen:2024nyc, Lee:2024esz}, understanding the dead-cone effect~\cite{Craft:2022kdo, Barata:2025uxp} studying jet modifications in the presence of a quark-gluon plasma~\cite{Andres:2022ovj, Yang:2023dwc, Bossi:2024qho, Barata:2023zqg, Barata:2023bhh, Andres:2023ymw, Singh:2024vwb, Singh:2025scb, Budhraja:2025ulx, Barata:2025fzd, Barata:2025zku}, and probing cold nuclear matter effects~\cite{Barata:2024wsu, Fu:2024pic, Devereaux:2023vjz}.

Generalizations to higher-point correlators, though valuable for probing multi-particle correlations and for, e.g.,~the determination of $\alpha_s$~\cite{CMS:2024mlf}, become increasingly cumbersome for large values of $N$ as they involve all $N \choose 2$ pairwise distances between the $N$ particles. While it is simpler to study projected $N$-point correlators (PENCs)~\cite{Chen:2020vvp}, obtained by integrating out the shape information and keeping only the largest separation between the $N$ particles, the computational cost remains as all these distances need to be compared to determine the largest separation. The projected correlators can also be extended to non-integer $N$ values through analytic continuation~\cite{Chen:2020vvp}, which provides a novel probe of the small-$x$ physics through jets in the $N\to 0$  limit. However, this analytic continuation scales even worse in terms of computational cost.\footnote{Indeed, in the original paper~\cite{Chen:2020vvp} only final states with up to four particles were considered.} As a result, despite the interesting physics reach, the computational challenge using the traditional parametrization hindered their practical use.  

In ref.~\cite{Alipour-fard:2024szj}, some of us developed a novel parametrization for $N$-point correlators that addresses this problem, reducing the computational cost for projected correlators to $\mathcal{O}(M^2 \ln M)$ \emph{independent} of $N$, thus allowing for their practical utilization. Our approach reorganizes the correlator by measuring the largest angular distance with respect to a ``special'' particle, thereby yielding a simpler geometrical structure. Importantly, in ref.~\cite{Alipour-fard:2024szj}, we showed that, in the collinear limit, this new parametrization provides a reasonable measure of the largest angular scale and preserves the analytic properties of energy correlators. In this paper, we will utilize this approach to study the non-perturbative corrections in the collinear regime and extend our analysis to the complementary back-to-back regime. 

The back-to-back limit of the EEC has been explored in the literature and analytic computations using the framework of soft collinear effective theory (SCET)~\cite{Bauer:2000yr,Bauer:2001yt,Bauer:2001ct,Bauer:2002nz} have reached the state-of-the art accuracy of next-to-next-to-next-to-next-to-leading-logarithmic (N$^4$LL) resummation~\cite{Duhr:2022yyp} including also for track-based EECs~\cite{Jaarsma:2025tck}.\footnote{ The back-to-back factorization for di-jet production was first established in ref.~\cite{Collins:1981uk}} Contrary to the collinear region, where only single logarithms of the angles appear, in the back-to-back regime, the observable receives double logarithmic (Sudakov) contributions due to the presence of soft and collinear radiation. Although the explicit energy weights still suppress the contribution of the soft radiation to the measured value of EEC at leading power, the soft radiation does contribute but only through a net recoil of the energetic jets. This simplicity allows the use of the well developed theoretical framework of $q_T$ resummation~\cite{Collins:1981va,Collins:1984kg} to obtain precise predictions for EEC in the back-to-back regime~\cite{deFlorian:2004mp, Tulipant:2017ybb, Kardos:2018kth, Moult:2018jzp, Ebert:2020sfi, Duhr:2022yyp, Aglietti:2024zhg, Aglietti:2024xwv,  Kang:2024dja,  Jaarsma:2025tck}. However, the extension to higher-point PENC distributions has so far not been explored.\footnote{For the top quark case, factorization for the three-point correlator in the back-to-back region was explored in~\cite{Pathak:2024}.} Such computations become highly complicated already for $N=4$ due to the complex phase-space structure for evaluating the projection. 

In this work, we develop the resummation framework for higher-point PENC distributions in the back-to-back region, providing all ingredients needed at next-to-next-to-leading-logarithmic (NNLL) accuracy. We will show how the simple phase-space appearing in our approach simplifies this framework. Our work is also motivated by phenomenological considerations:  First, the recent analysis of PENCs in the collinear limit, by the CMS collaboration has provided the most precise estimation of $\alpha_s$ from jet substructure measurements~\cite{CMS:2024mlf}. This analysis is constructed by considering the ratio of PE3C/EEC distributions which improves the precision of $\alpha_s$ determination~\cite{Chen:2023zlx} by suppressing major sources of systematics, though non-perturbative effects seem to be underestimated~\cite{Lee:2024esz}. A similar analysis in the back-to-back regime would complement this, reducing the dominant systematics, removing for example the relative fraction of quark- to gluon-jets that affects the collinear regime~\cite{Bossi:2024Nikhef}.

We derive the factorization formula for general PENC distributions within the framework of SCET, 
which in the back-to-back limit takes the canonical form as for any di-jet observable. This involves a hard function encoding the short-distance dynamics of the hard scattering, jet functions describing collinear radiation within each jet and encoding the measurement of PENCs, and a soft function capturing wide-angle soft emissions. Importantly, for the EEC observable and in general for PENCs, also for our new parametrization, the explicit energy weights ensure that soft radiation contributes only to the recoil between the collinear jets, in contrast to standard di-jet observables.\footnote{For standard di-jet observables, soft radiation directly contributes to the measurement. If, in addition, it also contributes to the recoil of the collinear partons, this significantly complicates the perturbative calculations. For instance, computations of jet broadening and in general, recoil-sensitive angularities are particularly challenging. The presence of recoil complicates the phase-space structure of radiation already at the next-to-leading-logarithmic order~\cite{Dokshitzer:1998kz, Becher:2011pf, Budhraja:2019mcz}.} Due to our new parametrization of PENCs, only one jet function is modified compared to the EEC ($N=2$) case. An important technical achievement of this work is then the calculation of this jet function at one-loop for arbitrary $N$, which was the only missing ingredient to achieve the desired NNLL accuracy\footnote{For an overview of the ingredients needed at different resummation orders, see e.g.~ref.~\cite{Almeida:2014uva}.} for general PENCs. The anomalous dimension of this jet function was already fixed by consistency to be equal to that of $q_T$ dependent observables. 

Next, we also utilize our approach to study the leading non-perturbative power corrections to arbitrary PENC distributions in the collinear regime. The direct contribution from soft radiation to the energy correlator is power-suppressed, but needs to be taken into account when considering the non-perturbative power corrections~\cite{Schindler:2023cww,Lee:2024esz,Chen:2024nyc,Budhraja:2024tev}. Furthermore, as highlighted in ref.~\cite{Lee:2024esz} for (integer) PENC distributions, these non-perturbative contributions play an important role in the phenomenological $\alpha_s$ determinations. In this study, we generalize the arguments presented in refs.~\cite{Lee:2024esz,Schindler:2023cww} to explore the effect of power corrections for general PENC distributions, characterizing their dependence on $N$, the center-of-mass energy and the angular distance. Employing our new parametrization, we find that the leading non-perturbative contributions are enhanced when $N \to 0$  up to the point that the non-perturbative corrections are no longer suppressed in angular scale compared to the perturbative contribution. They also have qualitatively different behavior, e.g., the power corrections no longer scale linearly with $N$~\cite{Budhraja:2024tev}. 

For $N>1$, it was recently shown~\cite{Lee:2024esz,Schindler:2023cww} that leading non-perturbative effects are captured by the parameter $\Omega_1$~\cite{Dokshitzer:1995zt,Dokshitzer:1997ew,Korchemsky:1999kt}.
For $N<1$, we find a new non-perturbative matrix element, that however can be related to $\Omega_1$ assuming it is dominated by a single non-perturbative gluon. Our estimates of the power corrections are compared to the default hadronization model used in {\textsc{Pythia}} simulations~\cite{Bierlich:2022pfr}, finding good agreement with our analytic estimates. 

The rest of the article is organized as follows: In sec.~\ref{sec:PENC}, we review the new parametrization we developed in ref.~\cite{Alipour-fard:2024szj} and extend it for the case of back-to-back jets in $e^+\, e^-$ collisions. Here, we also show the PE3C, comparing the traditional parametrization to our new approach, using di-jet  samples simulated by {\textsc{Pythia}}. In sec.~\ref{sec:b2b}, we derive the factorization theorem for PENCs in this regime and provide analytic results for the new one-loop jet function, showing explicitly the effect played by recoil on the computation of the jet function. In sec.~\ref{sec:nonp}, we investigate the size of leading non-perturbative contributions for PENCs and provide tests of our predictions against Monte Carlo simulations. We summarize in sec.~\ref{sec:summary}.

\section{New approach to PENCs}
\label{sec:PENC}

In the traditional parametrization of the projected $N$-point correlator, the shape information of the $N$-points is integrated out, keeping only the largest distance (angular separation) between all pairs of particles~\cite{Chen:2020vvp}. In ref.~\cite{Alipour-fard:2024szj}, some of us introduced a new parametrization that leads to a simpler phase-space structure, while maintaining the key physics properties in the collinear limit. There we showed that this enables a dramatic speed up, making the evaluation of the projected $N$-point correlator independent of $N$. We will recap the main points in this section and show in sec.~\ref{sec:b2b} that the simpler structure has important implications for the theoretical description of the back-to-back limit. In sec.~\ref{sec:nonp}, we will discuss the non-perturbative corrections in the collinear limit, using this parametrization.

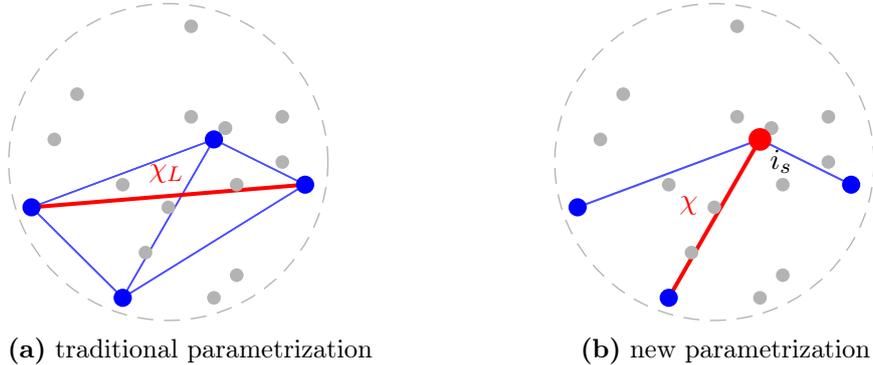
\begin{figure}
\centering
	\begin{tikzpicture}[scale=3]

\def\R{0.7}

\draw[
  dashed,
  dash pattern=on 6pt off 4pt,
  gray!60,
  line width=0.5pt
] (0,0) circle (\R);

\coordinate (P1) at ( 0.0, -0.2); 
\coordinate (P2) at ( 0.5, 0.2); 
\coordinate (P3) at (-0.4, 0.3); 
\coordinate (P4) at ( 0.3, -0.5); 
\coordinate (P5) at (-0.6, -0.2); 
\coordinate (P6) at ( 0.1, 0.6); 
\coordinate (P7) at (-0.2, -0.6); 
\coordinate (P8) at (-0.1, -0.4); 
\coordinate (P9) at ( 0.2, -0.6); 
\coordinate (P10) at ( 0.6, -0.1); 
\coordinate (P11) at (-0.5, 0.1); 
\coordinate (P12) at (0.1, 0.2); 
\coordinate (P13) at (0.5, 0.0); 
\coordinate (P14) at (0.25, 0.15); 
\coordinate (P15) at (-0.2, -0.1); 
\coordinate (P16) at (0.25, 0.15); 
\coordinate (P17) at (0.2, 0.1); 
\coordinate (P18) at (0.3, -0.1);

\def\Chosen{P17,P5,P7,P10}

\foreach \p in \Chosen {
  \foreach \q in \Chosen {
    \ifx\p\q\else
      \draw[blue!70, line width=0.6pt] (\p) -- (\q);
    \fi
  }
}

\draw[red, line width=1.5pt] (P5) -- (P10)
  node[pos=0.6, above left] {$\chi_L$};

\foreach \p in {P1,P2,P3,P4,P5,P6,P7,P8,P9,P10,P11,P12,P13,P14,P15,P16,P17,P18} {
  \fill[gray!60] (\p) circle (0.03);
}

\foreach \p in \Chosen {
  \fill[blue] (\p) circle (0.04);
}

\node[
  anchor=south west,
  xshift=-4pt,
  yshift=-20pt
] at (current bounding box.south west)
{\small\textbf{(a)}\ traditional parametrization};

\end{tikzpicture}\hspace{2cm}
    \begin{tikzpicture}[scale=3]

\def\R{0.7}

\draw[
  dashed,
  dash pattern=on 6pt off 4pt,
  gray!60,
  line width=0.5pt
] (0,0) circle (\R);

\coordinate (P1) at ( 0.0, -0.2); 
\coordinate (P2) at ( 0.5, 0.2); 
\coordinate (P3) at (-0.4, 0.3); 
\coordinate (P4) at ( 0.3, -0.5); 
\coordinate (P5) at (-0.6, -0.2); 
\coordinate (P6) at ( 0.1, 0.6); 
\coordinate (P7) at (-0.2, -0.6); 
\coordinate (P8) at (-0.1, -0.4); 
\coordinate (P9) at ( 0.2, -0.6); 
\coordinate (P10) at ( 0.6, -0.1); 
\coordinate (P11) at (-0.5, 0.1); 
\coordinate (P12) at (0.1, 0.2); 
\coordinate (P13) at (0.5, 0.0); 
\coordinate (P14) at (0.25, 0.15); 
\coordinate (P15) at (-0.2, -0.1); 
\coordinate (P16) at (0.25, 0.15); 
\coordinate (P17) at (0.2, 0.1); 
\coordinate (P18) at (0.3, -0.1);

\def\Chosen{P17,P5,P7,P10}

\def\Special{P17}

\foreach \p in \Chosen {
  \ifx\p\Special\else
    \draw[blue!70, line width=0.8pt] (\Special) -- (\p);
  \fi
}

\draw[red, line width=1.5pt] (P17) -- (P7)
  node[pos=0.55, above left] {$\chi$};

\foreach \p in {P1,P2,P3,P4,P5,P6,P7,P8,P9,P10,P11,P12,P13,P14,P15,P16,P17,P18} {
  \fill[gray!60] (\p) circle (0.03);
}

\foreach \p in \Chosen {
  \fill[blue] (\p) circle (0.04);
}

\fill[red] (\Special) circle (0.05);

\node[below right] at (P17) {$i_s$};

\node[
  anchor=south west,
  xshift=6pt,
  yshift=-20pt
] at (current bounding box.south west)
{\small\textbf{(b)}\ new parametrization};

\end{tikzpicture}
    \caption{
        An illustrative diagram comparing the (a) traditional parametrization for PENCs against the (b) new parametrization for $N=4$.
        In (a) all \(\binom{N}{2}\) pairwise distances are needed to determine the largest separation $\chi_L =\max_{j,k}\{\theta_{i_k,i_j}\}$. By contrast, in (b) the PENCs are parametrized relative to the special particle $i_s$, with the largest separation characterized by $\chi =\max_{j}\{\theta_{i_s,i_j}\}$. In both cases all particles are summed over, including the special particle in (b). The dashed line illustrates the boundary of the jet, which is not needed in $e^+e^-$ collisions.
    }
	\label{fig:cartoon}
\end{figure}

The key idea in our approach is to consider only the distances of all particles relative to a ``special'' particle $i_s$, and subsequently perform an energy-weighted sum over all possible choices of $i_s$. A cartoon illustrating the difference between the traditional parametrization and our new parametrization is shown in fig.~\ref{fig:cartoon}.

Concretely, the $N$-point projected energy correlator is defined as 
\begin{align}
    \label{eq:new_def}
    \frac{1}{\sigma}\frac{\df \sigma^{[N]}}{\df \chi}
    &= 
    \frac{1}{{\sigma}}
    \int \df\sigma_X
        \sum_{i_s}  z_{i_s} \!\!
        \sum_{i_1 \dots i_{N-1}}
       \!\! z_{i_1} \dots z_{i_{N-1}} 
        \delta\bigl(
            \chi 
            \!-\! 
            \max_j\{\theta_{i_s,i_j}\}
        \bigr)
    .
\end{align}
Here $\df\sigma_X$ denotes the differential cross section for producing a hadronic final-state $X$, and the sums over $i_s$ and $\{i_j\}_{j=1}^{N-1}$ run over all $M$ particles in $X$, $z_i = 2E_i/Q$ are the energy fractions with respect to half the center-of-mass energy of the collision.\footnote{This extends the jet-based parametrization in ref.~\cite{Alipour-fard:2024szj} to a full $e^+\, e^-$ event shape. We divide by half the center-of-mass energy in $z_i$, as this corresponds to the jet energy in the dijet limit.} The angular separation of particle $i_j$ from the special particle $i_s$ is denoted by $\theta_{i_{s},i_{j}}$ while $\chi$ corresponds to the largest (maximum) angular separation of the $N-1$ final-state particles $i_j$ relative to $i_s$. For the special case of $N=2$ the traditional and new parametrization coincide.

We will also use the cumulative distribution, 
\begin{align}
    \label{eq:new_def_cumul}
    \Sigma^{[N]}(\chi)
    &= 
    \int_0^{{\chi}}\! \df \chi'\, \frac{\df \sigma^{[N]}}{\df \chi'}
    =
    \int\! \df \sigma_X
        \sum_{i_s=1}^M  z_{i_s} \bigl[
        z_{\rm disk}(i_s,\chi)\bigr]^{N-1}\! ,
\end{align}
which provides a simple interpretation of our approach in terms of the total energy fraction $z_{\rm disk}(i_s,\chi)$ in a disk of radius $\chi$ around the special particle $i_s$.
This expression emphasizes that, unlike the traditional parametrization, no complications arise when  generalizing to non-integer values of $N$.
It also shows that these PENCs obey the sum rule 
\begin{equation}
    \frac{1}{\sigma} \Sigma^{[N]}({\pi}) = {2^N}\, ,
\end{equation}
which follows directly from energy conservation as $z_{\rm disk}{(i_s,\pi)} = {2}$, and the summing over all possible choices of the special particle also gives $\sum_{i_s} z_{i_s} = {2}$. 

The cumulative distribution in eq.~\eqref{eq:new_def_cumul} allows us to highlight the major computational advantage of this new parametrization in reducing the scaling in time of the $N$-point projected correlator for $M$ particles from $\mathcal{O}(M^N)$ to a simple $\mathcal{O}(M^2 \ln M)$ scaling, independent of the value of $N$. 
One factor of $M$ arises from the sum over $i_s$, while the  $M \ln M$ comes from sorting the particles by their distance relative to $i_s$, as explained in ref.~\cite{Alipour-fard:2024szj}. 

\begin{figure}
\centering

\begin{subfigure}{0.49\textwidth}
    \centering
    \includegraphics[width=\linewidth]{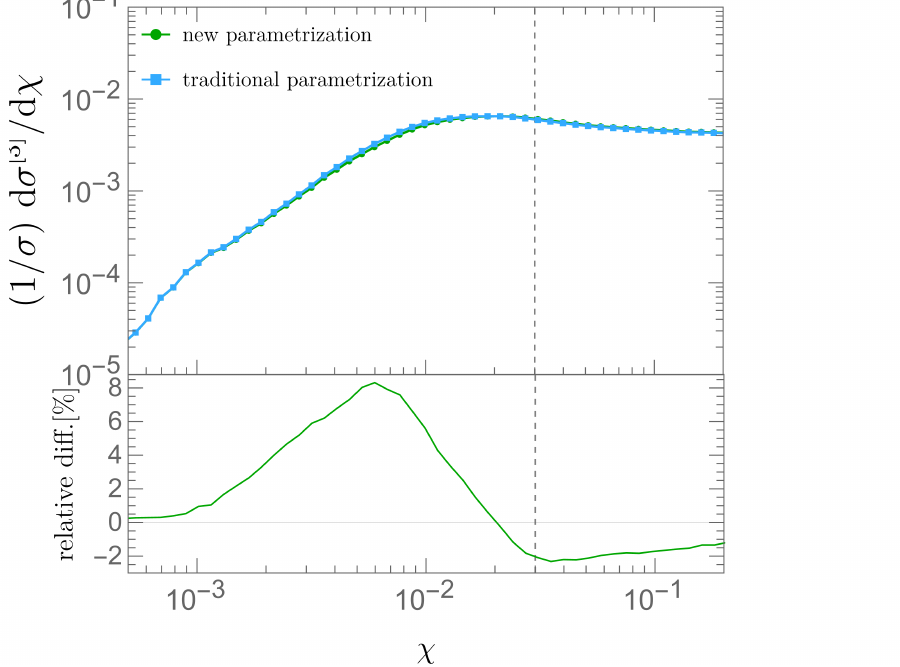}
    \caption{}
\end{subfigure}
\hfill
\begin{subfigure}{0.49\textwidth}
    \centering
    \includegraphics[width=\linewidth]{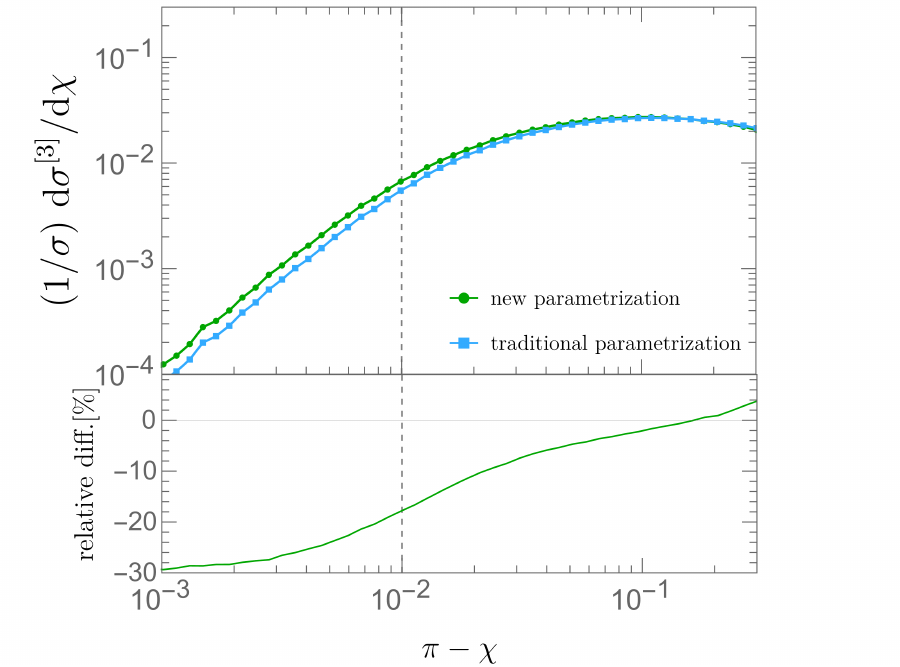}
    \caption{}
\end{subfigure}

\caption{
The PE3C distribution in $e^{+}\, e^{-} \to {\rm jets}$, obtained from \textsc{Pythia} at $Q=1$ TeV, for the traditional parametrization and our new parametrization logarithmic in (a) $\chi$ and (b) $\pi - \chi$. The relative difference is shown in the bottom panel of the figure. The vertical dashed line indicates the transition between the perturbative and non-perturbative region, and the plot range is chosen to highlight the collinear region in (a) and the back-to-back regime in (b). As visible from the figure, the differences in the perturbative region are relatively small: for (a) a few percent and up to about 10\% for (b), while difference in the transition to the non-perturbative region are larger but below $10\%$ in (a) and up to $30 \%$ in (b). The larger differences for the back-to-back regime are explained in the text. 
}
\label{fig:pe3c}
\end{figure}

On the theoretical side, this new parametrization has the same scaling behavior in the collinear limit, $\chi \to 0$, as the traditional parametrization. In the perturbative region it is given by 
\begin{equation} \label{eq:pert_coll_scaling}
    \frac{\df\sigma^{[N]}}{\df\chi} \sim \chi^{\gamma(N+1)-1}\, ,
\end{equation}
with differences with the traditional parametrization starting at next-to-next-to-leading logarithmic (NNLL) accuracy~\cite{Alipour-fard:2024szj}. This translates to percent-level differences in the perturbative scaling regime for $N=3$, as shown in the comparison between our parametrization and the traditional parametrization in fig.~\ref{fig:pe3c}, using simulated $e^+e^-$ collisions generated with {\textsc{Pythia} 8.3} at $Q = 1\, {\rm TeV}$~\cite{Bierlich:2022pfr}.

The scaling in eq.~\eqref{eq:pert_coll_scaling} is governed by the anomalous dimensions of the underlying theory  $\gamma(N\!+\!1)$~\cite{Chen:2020vvp}, which can be expressed as the $N$th-moment of the time-like splitting functions $P(x)$.\footnote{Technically, it should be the eigenvalues of these matrices, as the splitting functions include mixing between quarks and gluons.} 
In the deeply non-perturbative region of the collinear limit the scaling is the same for the traditional and new parametrization, and corresponds to a free hadron gas. The largest differences arise in the less-understood transition region between perturbative and non-perturbative scaling, which for $N=3$ are $\lesssim$ 10\%, see fig.~\ref{fig:pe3c}(a). 

In the derivation of eq.~\eqref{eq:pert_coll_scaling}, the inequality $\chi \leq \chi_L \leq 2 \chi$ plays an important role, where $\chi_L$ would be the corresponding angle for the traditional parametrization~\cite{Alipour-fard:2024szj}. 
Roughly speaking, in the collinear limit the logarithm of $\chi$ is the natural variable, due to the power-law scaling, and the difference of at most a factor of 2 does not affect that scaling. However, in the fixed-order (or bulk) region between the collinear and back-to-back limits,  where there is no power-law scaling, this difference of up to a factor of 2 is relevant, leading to larger differences between the energy correlator for the traditional and new parametrization.

In the next section we will focus on the back-to-back limit, $\chi \to \pi$, deriving the factorization. In this limit the cross section contains double logarithms of $\pi - \chi$, because it becomes sensitive to the recoil of soft radiation~\cite{Tulipant:2017ybb, Kardos:2018kth, deFlorian:2004mp, Aglietti:2024zhg, Aglietti:2024xwv, Ebert:2020sfi, Kang:2024dja, Moult:2018jzp, Duhr:2022yyp, Jaarsma:2025tck}.
Though we will also reach the conclusions that in this limit the differences between the traditional and new parametrization start at NNLL, numerically these differences will be much larger because the cross section contains double instead of single logarithms. 
Indeed, for $N=3$ we find differences of up to 10\% in the perturbative back-to-back region, as shown in fig.~\ref{fig:pe3c}(b), while  differences in the perturbative collinear region still remain at the few percent level.

Importantly, the simpler phase-space structure resulting from the new parametrization makes it straightforward to generalize the theoretical description of the 2-point correlator to the projected $N$-point correlator in the yet unexplored back-to-back limit, and we will provide the ingredients necessary for NNLL resummation.

\section{PENCs in the back-to-back limit} 
\label{sec:b2b}

In this section, we provide analytic results for the new  PENCs in the back-to-back limit. We start by discussing in detail the kinematics of PENCs in sec.~\ref{sec:b2b_kin}, and present the factorization in this regime in sec.~\ref{sec:b2b_fact}. We provide explicit one-loop results for the new jet function in sec.~\ref{sec:b2b_jet}, which is the only extra ingredient needed to achieve NNLL accuracy.

\subsection{Kinematics of PENCs in the back-to-back limit}
\label{sec:b2b_kin}

The structure of the two-point energy correlator has been investigated extensively in the literature in the back-to-back region~\cite{deFlorian:2004mp, Tulipant:2017ybb, Kardos:2018kth, Moult:2018jzp, Ebert:2020sfi, Duhr:2022yyp, Aglietti:2024zhg, Aglietti:2024xwv,  Kang:2024dja,  Jaarsma:2025tck}. Here, we briefly review it, allowing us to introduce the necessary concepts and notation, after which we extend to PENCs with general $N$.

For $e^+\, e^-$ collisions, it is convenient to use
\begin{equation}\label{eq:x}
    x = \frac{1-\cos\chi}{2} \, .
\end{equation}
instead of the angle $\chi$, such that the back-to-back limit $\chi \to \pi$ corresponds to $x \to 1$. In this limit, the event consists of two nearly back-to-back jets, which will be described by collinear radiation. The offset from exact back-to-back jets is due to both the collinear and soft radiation; with soft radiation providing a color connection between the two jets. The energy weights in the definition of the EEC (see eq.~\eqref{eq:new_def} with $N=2$) suppress any direct contribution from soft radiation at leading power, such that its only  effect is through recoil. This will no longer be the case for the sub-leading corrections due to hadronization in sec.~\ref{sec:nonp}.

We will use light-cone variables to parametrize the momenta of the particles in the final state. In these coordinates, any four-momentum can be decomposed as 
\begin{align}
p^\mu = n\cdot p \frac{\bar{n}^\mu}{2} + \bar{n}\cdot p \frac{n^\mu}{2} + {\bf p}
^\mu = (n \cdot p, \bar n \cdot p, {\bf p}^\mu)
\end{align}
where $n = (1,0,0,1)$ and $\bar{n} = (1,0,0,-1)$ are the light-like directions along the initial quark and anti-quark produced in the $e^+e^-$ collision.\footnote{While this specific frame is not experimentally accessible, it simplifies the discussion, and the final result is frame independent.} We omit the $\perp$ on the transverse components ${\bf p}^\mu$ for brevity.

In terms of the effective theory formulation, the relevant modes that contribute to the EEC in the back-to-back limit are: two collinear modes describing the energetic radiation in the two jets,  and a soft mode that encodes the recoil of these jets due to the low energy radiation. These modes are described by the following scaling of their light-cone momenta,
\begin{align}
\label{eq:pc}
n\text{-collinear:} & & p_n &\sim Q(1,\lambda^2,\lambda)\, ,  \hspace{15ex} \nn \\
\bar{n}\text{-collinear:} & & p_{\bar{n}} &\sim Q(\lambda^2,1,\lambda) \,, \nn \\
\text{soft:} & & p_{\rm soft} &\sim Q(\lambda,\lambda,\lambda)\,, \vphantom{\lambda^2}
\end{align}
in terms of the power counting parameter $\lambda \sim \sqrt{1-x} \sim \pi - \chi$ and the center-of-mass energy $Q$ of the collision. The effective theory whose modes satisfy the power counting in eq.~\eqref{eq:pc} is known as SCET$_{\rm II}$~\cite{Bauer:2002aj}.
As the recoil concerns the momentum component perpendicular to the jet, the corresponding ${\bf p}$ component has the same $\lambda$-scaling for all these modes. 

Indeed, the expression of the angular scale $x$ for the $n$-collinear parton $i$ and $\bar n$-collinear parton $j$ is~\cite{Moult:2018jzp}
\begin{align} \label{eq:kin}
    1-x = \frac{{\bf p}^2}{Q^2} &= \frac{1}{Q^2}\,\Big( \frac{{\bf p}_{i}'}{z_i} + \frac{{\bf p}_{j}'}{z_j} - {\bf p}_{{\rm soft}} \Big)^2 
    \nn \\ 
    &= \frac{1}{Q^2}\,({\bf p}_{i} + {\bf p}_{j} - {\bf p}_{{\rm soft}})^2 
    \, ,
\end{align}
where we have dropped power corrections. Here, $z_{i,j}$ denote the fraction of the jet energy carried by partons $i$ and $j$, ${\bf p}_{i,j}'$ their transverse momenta with respect to the initiating parton (or field), and ${\bf p}_{{\rm soft}}$ is the \emph{total} transverse momentum of soft radiation. Since only this specific ratio of the hadronic variables ${\bf p}_{i,j}'$ and $z_{i,j}$ appears, it is convenient to switch to their ratio, which is a partonic variable that we denote by ${\bf p}_{i,j}$.

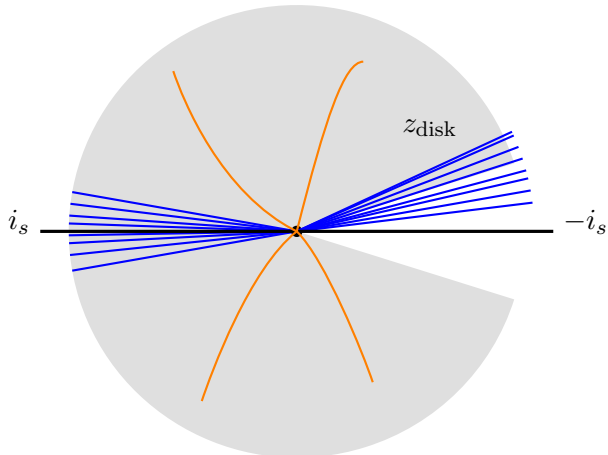
\begin{figure}
\centering
	\begin{tikzpicture}[scale=2.5]
	
	\coordinate (O) at (0,0);

    \def\R{1.2}      
    \def\cut{35}      

    \fill[gray!25]
    ({\cut/2}:\R) 
    arc ({\cut/2}:{360-\cut/2}:\R) 
    -- (0,0) -- cycle;
	
	\def\phi{15}
	
	\fill (O) circle (0.03);
	
	\begin{scope}[rotate around={\phi:(O)}]
		
		\foreach \ang in {-8,-5,-2,0,3,6,9,10}
		{
			\draw[blue, thick] (O) -- ++({1.25*cos(\ang)},{1.25*sin(\ang)});
		}
		
	\end{scope}
	
	\foreach \ang in {-10,-6,-3,-1,2,4,7,10}
	{
		\draw[blue, thick] (O) -- ++({-1.2*cos(\ang)},{1.2*sin(\ang)});
	}


	\draw[very thick] (-1.35,0) -- (1.35,0);
	
	\node[left] at (-1.35,0.05) {$i_s$};
	\node[right] at (1.35,0.05) {$-i_s$};
		

\draw[orange, thick]
(0,0) .. controls (0.15,0.6) and (0.25,0.9) .. (0.35,0.9);

\draw[orange, thick]
(0,0) .. controls (-0.25,-0.2) and (-0.45,-0.75) .. (-0.5,-0.9);

\draw[orange, thick]
(0,0) .. controls (0.15,-0.15) and (0.35,-0.65) .. (0.4,-0.8);

\draw[orange, thick]
(0,0) .. controls (-0.4,0.2) and (-0.6,0.7) .. (-0.65,0.85);

\node[black] at (0.7,0.55) {$z_{\text{disk}}$};
	
\end{tikzpicture}
    \caption{
        Schematic of the back-to-back regime, where the special particle is in  one of the jets and the measurement is sensitive to details of the collinear radiation in the opposite jet, through their distance to the antipode of the special particle $-i_s$, i.e.~whether these particles are included in $z_{\rm disk}$ shown in gray. The soft radiation produces an overall recoil, such that the jets are no longer exactly back-to-back. 
    }
	\label{fig:b2b}
\end{figure}

For the traditional parametrization, generalizing these arguments to PENCs is non-trivial because all possible distances between subsets of particles in \emph{both} jets are needed to compute their largest separation. In our approach, the phase-space structure is simplified because we only have to consider distances to the special particle. If the special particle is in one jet, we just have to consider the distances to subsets of particles in the other jet. Instead of the largest distance $\chi$ we can equivalently consider the smallest distance $\pi-\chi$ to the antipode of the special particle, depicted in fig.~\ref{fig:b2b}. This means that the complications for $N\neq 2$ reside entirely in the other jet function, while the jet containing the special particle is described by the same jet function $J$ as for $N=2$. Indeed, collinear emissions in the jet containing the special particle are always included in the disk of radius $\chi$ when $\chi \to \pi$, i.e.~they never set the largest distance and the sum over their momentum fractions is 1.\footnote{We find it convenient to use a convention in which the total momentum fraction of all collinear radiation in a jet is 1, because otherwise the number of detectors in the jet with the special particle would yield a nontrivial overall factor for the jet function.} 

The new jet function, which we denote as $J_{N-1}$, takes as input the antipode of the special particle, characterized by ${\bf p}_{{\rm ap}} = -{\bf p}_{s}'/z_s + {\bf p}_{{\rm soft}} = -{\bf p}_{s} + {\bf p}_{{\rm soft}}$ where $s$ labels the special particle (which is not soft). The term ${\bf p}_{{\rm soft}}$ accounts for the recoil from soft radiation. This complicates the factorization structure compared to the simpler additive structure in eq.~\eqref{eq:kin}, where the jet function's contributions to the distance $x$ can be calculated independently of each other. Because the jet is insensitive to the initiating quark spin at leading power, the jet function $J_{N-1}$ has azimuthal symmetry, it can only depend on the \emph{relative} angle $\phi$ between its contribution to the transverse momentum ${\bf p}$ and ${\bf p}_{ \rm ap}$.
At leading order, the jet function contains a single parton, which is automatically the one closest to the antipode, and the measurement can still be written as in \eqref{eq:kin}.

With the knowledge of the modes that contribute in the back-to-back regime and a discussion of the observable, we are now ready to write down the factorization theorem and discuss its ingredients, in the following section.

\subsection{Factorization in the back-to-back limit}
\label{sec:b2b_fact}

The factorized cross-section for PENCs can be written as a product of the hard function $H$, that characterizes the hard scattering event generating the energetic back-to-back partons $e^+\, e^- \to q {\bar q}$, with a convolution between the jet function $J$ and the soft function $S$ governing the evolution of the energetic quark and anti-quark at subsequent energy scales.\footnote{As we are considering $e^+\, e^-$ collisions, we do not have to consider possible Glauber exchanges between incoming hadrons.} In particular, the jet function $J$ describes the dynamics of collinear radiation inside the jet, while the soft function $S$ accounts for the low energy exchanges between the two back-to-back jets. At leading power in SCET, the interactions between the jet and soft degrees of freedom are decoupled through a BPS field redefinition~\cite{Bauer:2001yt} and only appear in the current.\footnote{Although this is strictly a SCET$_{\rm I}$ statement, it is common to match onto SCET$_{\rm II}$ through an intermediate SCET$_{\rm I}$ theory, which shows how Wilson lines appear.} As a result, there are no interactions between collinear and soft radiation, and each of the hard, jet and soft functions describe the dynamics at a single scale. This allows the large logarithms of $1-x$ to be resummed through the solution of the renormalization group equations, evolving the respective functions from their natural scale to a single common scale.

In terms of these ingredients, the factorized cross-section in the back-to-back regime can be written as 
\begin{align}
\label{eq:factorization}
    \frac{1}{\sigma_0}\frac{\df\sigma^{[N]}}{\df x} &= H(Q;\mu) \int \df^2 {\bf p}_{n}\,  J({\bf p}_{n}^2;\mu,\nu)     
    \int \df^2 {\bf p}_{\text{soft}}\, S({\bf p}_{\text{soft}}^2;\mu,\nu)
    \int \df^2 {\bf p}_{{\bar n}}\,   \\
    &\quad  \times J_{N-1}({\bf p}_{{\bar n}},-{\bf p}_{n} + {\bf p}_{\text{soft}};\mu,\nu) \,\,
    \delta\Bigl[1-x-\frac{1}{Q^2} \bigl({\bf p}_{n} + {\bf p}_{{\bar n}} - {\bf p}_{\text{soft}}\bigr)^2 \Bigr] +  \{n \leftrightarrow {\bar n} \nn \}\, .
\end{align}       
The measurement $x$ of the largest distance is fixed in terms of the contributions from particles in the jet and soft sectors, described by the transverse momentum variables ${\bf p}_{\{n,\bar n,{\text{soft}}\}}$. 
Here $J$ is the same jet function as for the EEC ($N=2$)~\cite{Moult:2018jzp} and does not involve any $N$-dependence because in the back-to-back regime the particles in the jet with the special particle are always included in $z_{\rm disk}$.  
The other jet function, describing the jet without the special particle, and containing (up to) $N-1$ detectors, is denoted by $J_{N-1}$. $J_{N-1}$ contains all multinomial factors, arising for higher-point correlators, associated with how many of the $N$ detectors are on the jet~\cite{Chen:2020vvp,Budhraja:2024xiq,Budhraja:2024tev}.
Finally, $\mu$ ($\nu$) represent the renormalization scales for (rapidity) renormalization group evolution (RGE). We will use the rapidity regulator of ref.~\cite{Becher:2011dz}, for other choices of the rapidity regulator, see e.g., refs.~\cite{Collins:2011zzd, Ji:2004wu, Chiu:2009yx, Chiu:2011qc, Li:2016axz}. The presence of rapidity divergences and the corresponding $\nu$-evolution is intrinsic of the fact that PENCs in the back-to-back limit are SCET$_{\rm II}$ observables. 

The hard function $H$ in the above equation is the same as that for any $e^+\, e^- \to $ di-jet observable. At next-to-leading order (NLO), the hard function is given by~\cite{Manohar:2003vb,Bauer:2003di} 
\begin{align}
    H(Q; \mu) &= 1+\frac{\alpha_s(\mu)C_F}{\pi} \Bigl(-2\ln^2\frac{\mu}{Q}-3\ln\frac{\mu}{Q}-4+\frac{7}{2}\zeta_2 \Bigr) \, ,
\end{align}
The soft function $S$ only contributes to the observable through an overall recoil momentum $\bf{p}_{\text{soft}}$. Consequently, it remains the same as  for the two-point energy correlator. The soft function has been evaluated up to three loops and the corresponding results (including the one-loop) can be found in refs.~\cite{Moult:2018jzp, Jaarsma:2025tck}. The resummation is most easily carried out in the Fourier conjugate space, for which the soft function up to NLO is given by 
\begin{align}
    S({\bf b}^2;\mu,\nu) &= 1 + \frac{\alpha_s(\mu)C_F}{\pi} \biggl[-\frac{1}{2}\ln^2\Bigl(\frac{{\bf b}^2 \mu^2 e^{2\gamma_E}}{4}\Bigr) + \ln\Bigl(\frac{{\bf b}^2 \mu^2 e^{2\gamma_E}}{4}\Bigr) \ln\Bigl(\frac{\mu}{\nu}\Bigr) -2\zeta_2 \biggr]\, 
\,.\end{align}
Here the impact parameter ${\bf b}$ is Fourier conjugate to ${\bf p}_{\text{soft}}$.

The jet function $J = J_1$, i.e.~$J_{N-1}$ with $N=2$. Consequently, we don't discuss it separately, but leave it for the next section in which we discuss the calculation of $J_{N-1}$. Nevertheless, $J_1$ is substantially simpler, and has been obtained by taking moments of the transverse momentum dependent fragmentation functions~\cite{Collins:2011zzd,Echevarria:2016scs}. Only the jet function $J_{N-1}$ is sensitive to the $N$-dependence of the measured $N$-point correlator.  
At NLL accuracy, where the resummation only requires the tree-level hard, jet, and soft functions and the evolution is determined by their anomalous dimensions, there is no difference between the traditional and new parametrizations. In particular, the tree-level jet functions each involve a single collinear parton, for which this difference is irrelevant.
As highlighted in the previous section, the simplicity of our approach allows us to generalize the computation to any arbitrary PENC without the requirement of additional phase-space considerations. 

\subsection{One-loop jet function for PENC}
\label{sec:b2b_jet}

We now compute the new jet function $J_{N-1}$ for the jet not containing the special particle, at one-loop accuracy. This contains all non-trivial $N$-dependence. As previously stated, the jet function $J$ with the special particle is equal to the jet function for the two-point energy correlator, $J = J_1$. 

The jet function can be computed from the QCD splitting functions, as was shown at NNLO in ref.~\cite{Ritzmann:2014mka}. This averts the use of the more complicated collinear Feynman rules in SCET. Writing the (bare) jet function in terms of a collinear phase-space integral over the $P_{qq}$ QCD splitting function,\footnote{Technically, dimensional regularization yields a $(\sin \phi')^{-2\epsilon}$ dependence on the angle $\phi'$ of ${\bf k}$ in the transverse plane. In our case this does not affect the result.}
\begin{align} \label{eq:J_NLO}
    J_{N-1}({\bf p}, {\bf p}_{\text{ap}};\mu,\nu) &= \frac{\alpha_s C_F}{2\pi^2} \frac{e^{\epsilon \gamma_E}\mu^{2\epsilon}}{\Gamma(1\!-\!\epsilon)} \Big(\frac{\nu}{Q}\Big)^\eta 
    \int\! \frac{\df z}{(1\!-\!z)^{\eta}} \int\! \frac{\df^2 {\bf k}}{({\bf k}^2)^{1+\epsilon}} P_{qq}(z)\, \mathcal{O}_{N-1}({\bf p},{\bf p}_{\text{ap}},z,{\bf k}) \, , \nn \\
    P_{qq}(z) &= \frac{1+z^2}{1-z} - \epsilon(1-z)\,.
\end{align}
Here $\epsilon$ ($\eta$) is the dimensional (rapidity) regulator, with the associated (rapidity) renormalization scale denoted by $\mu$ ($\nu$). The strong coupling is denoted by $\alpha_s$,
and $C_F = 4/3$ is the color factor for quark-initiated jets. Finally,
$\mathcal{O}$ encodes the measurement, which in the absence of recoil ${\bf p}_{\text{ap}} = {\bf 0}$ is given by
\begin{align}
    \mathcal{O}_{N-1}({\bf p},{\bf 0},z,{\bf k}) &=  \Theta\Bigl(z<\frac12\Bigr) \biggr\{[(1+z)^{N-1} - 1^{N-1}]\, \delta^2\Bigl({\bf p} - \frac{{\bf k}}{z}\Bigr) \nn \\
    &\quad + [2^{N-1} - (1\!+\!z)^{N-1}]\, \delta^2\Bigl({\bf p} + \frac{{\bf k}}{1-z}\Bigr)\biggr\} + (z \leftrightarrow 1\!-\!z)\,. 
\end{align}
The $\Theta$-function encodes that if $z<\tfrac12$, the particle with momentum fraction $1-z$ is closer to the antipodal point and therefore gets added later to $z_{\rm disk}$. The change in $z_{\rm disk}$, shown in square brackets, contributes at the distance of the corresponding emission, encoded in ${\bf p}$.
Performing the $z$ and ${\bf k}$ integral, renormalizing the expression, and including the tree-level result
\begin{align}
  J_{N-1}({\bf p}, {\bf 0}; \mu,\nu) &= (2^{N-1}-1) \biggl(\delta^2({\bf p}) + \frac{\alpha_s C_F}{\pi} \biggl\{ j_{N-1} \delta^2({\bf p})
  \!+\! \biggl[2 \ln \Bigl(\frac{Q}{\nu}\Bigr) \!-\! \frac32 \biggr] \mathcal{L}_0^T({\bf p},\mu)  \biggr\}\biggr)\,,
\end{align}
where we find it convenient to pull out the tree-level coefficient $2^{N-1}-1$ and $\mathcal{L}_0^T$ is the plus distribution defined through
\begin{equation}
    \mathcal{L}_n^T({\bf p},\mu) = \frac{1}{2\pi\mu^2}\Bigl[\frac{\ln^n({\bf p}^2/\mu^2)}{{\bf p}^2/\mu^2}\Bigr]_{+}
\,.\end{equation}
The general form for the coefficient $j_{N-1}$ of $\delta^2({\bf p})$ is
\begin{align}
    j_{N-1} &= \frac{1}{2^{N-1}-1} \int \df z\, \biggl( \frac{2^{N-1}-1}{2} (1-z) \,+ \frac{1+z^2}{1-z}\biggl\{ \Big[((1+z)^{N-1}-1)\, \ln z \nn \\
    &\quad  +\, (2^{N-1}-(1+z)^{N-1}) \, \ln (1-z)\Bigr] \Theta(z<1/2) + (z \leftrightarrow 1-z)
    \biggr\} \biggr) \, , 
\end{align}
which for specific values of $N$ results in
\begin{equation}
  j_{-0.5} = -3.04 \,, \quad j_1 = 1- 2 \zeta_2\, ,\quad j_2 = \frac{25}{24} + \frac12 \ln 2 - 2 \zeta_2 \,, \quad
  j_3 = \frac{57}{56} + \frac{27}{28} \ln 2 - 2 \zeta_2 \,.
\end{equation}

We now move on to the case of recoil, which we write as 
\begin{equation}
J_{N-1}({\bf p}, {\bf p}_{\text{ap}};\mu,\nu) = J_{N-1}({\bf p}, {\bf 0};\mu,\nu) + \Delta J_{N-1}({\bf p}, {\bf p}_{\text{ap}};\mu,\nu)
\,.\end{equation}
The expression for $\Delta J$ is the same as eq.~\eqref{eq:J_NLO}, with the measurement $\mathcal O$ replace by $\Delta {\mathcal O}$, given in eq.~\eqref{eq:Delta_J} below. Moreover, as the recoil correction is finite, there
is no dependence on the scales $\mu,\nu$ in $\Delta J$ at this order.
Compared to the case without recoil, the boundary is no longer simply at $z=\tfrac12$ but follows from
\begin{equation}
    \Bigl|\frac{{\bf k}}{z} - {\bf p}_{ \text{ap}} \Bigr|
    = \Bigl|\frac{{\bf k}}{1-z} + {\bf p}_{ \text{ap}} \Bigr|
\,,\end{equation}
which has as solution
\begin{equation}
    z_*' = \frac12 + \frac{|{\bf k}| - \sqrt{{\bf k}^2+ {\bf p}_{\text{ap}}^2 \cos^2 \phi'}}{2 |{\bf p}_{\text{ap}}| \cos \phi'}\,,
\end{equation}
where $\phi'$ is the angle between ${\bf k}$ and ${\bf p}_{\text{ap}}$. For $0 < \phi' < \tfrac{\pi}{2}$ this satisfies $z_*' < \tfrac12$, while for $\tfrac{\pi}{2} < \phi' < \pi$ instead $z_*' > \tfrac12$. 

However, this form is not so convenient, because there is a delta function that will cause ${\bf k}$ to be replaced by $z {\bf p}$ or $-(1-z) {\bf p}$. 
To determine the boundary in terms of ${\bf p}$, we consider these two cases separately, 
\begin{align} \label{eq:z_star}
    z_* = 
    \begin{cases}
    \dfrac12 - \dfrac{|{\bf p}_{\text{ap}}| \cos \phi}{2 (|{\bf p}| - |{\bf p}_{\text{ap}}| \cos \phi)} & \cos \phi < \dfrac{|{\bf p}|}{2|{\bf p}_{\text{ap}}|}\,, \\
       0 & \cos \phi > \dfrac{|{\bf p}|}{2|{\bf p}_{\text{ap}}|}\,,
    \end{cases}
\end{align}
where $z_*$ (or equivalently, $1-z_*$) corresponds to the case where the particle with momentum fraction $z$ (or, $1-z$) sets ${\bf p}$. 
Note that $\phi$ now denotes the angle between ${\bf p}$ and ${\bf p}_{\text{ap}}$, though again $z_* < \tfrac12$ for $0 < \phi < \tfrac{\pi}{2}$ and $z_* > \tfrac12$ for $\tfrac{\pi}{2} < \phi < \pi$.
We thus need to correct the no-recoil measurement as follows
\begin{align}
    \Delta \mathcal{O}_{N-1}({\bf p},{\bf p}_{\text{ap}},z,{\bf k}) &= \Bigl\{[2^{N-1} - (2\!-\!z)^{N-1}] - [(1+z)^{N-1} - 1^{N-1}] \Bigr\} \\ \nn
    &\quad \times
     \biggl[\Theta\Bigl(z_* \!<\!z\!<\!\frac12\Bigr) - \Theta\Bigl(\frac12 \!<\! z \!<\! z_* \Bigr)\biggr] \delta^2\Bigl({\bf p} - \frac{{\bf k}}{z}\Bigr) + (z \leftrightarrow 1-z)\,.
\end{align}
Note that for $N=2$, $\Delta {\mathcal O}_1 = 0$ and thus $\Delta J_1 =0$, as expected. The $1/(1-z)$ soft divergence of the splitting function will be regulated by this measurement, because $z_* <1$, while when $z_* = 0$ (for the $z \leftrightarrow 1\!-\!z$ term)
\begin{align}
[2^{N-1} \!-\! (2 \!- \!z)^{N-1}] \!-\! [(1\!+\!z)^{N-1} \!-\! 1] = \mathcal{O}(z)
\,,\end{align}
thus taming the singularity.

For general $N$, this leads to a finite correction $\Delta J_{N-1}$, which can be written as\footnote{While the expression contains explicit $\mu$-dependence, this dependence cancels in the final result.} 
\begin{align} \label{eq:Delta_J}
    \Delta J_{N-1}(|{\bf p}|,|{\bf p}_{\text{ap}}|, \phi) &= \frac{\alpha_s C_F}{\pi} 
   \biggl[ \int\! \df^2 {\bf k}\,\df z\, \mathcal{L}_0^T({\bf k},\mu)\, \frac{1+z^2}{1-z}\, 
    \Delta \mathcal{O}_{N-1}({\bf p},{\bf p}_{\text{ap}},z,{\bf k})   \\
    &\quad\quad\quad\quad  + \delta^2({\bf p})\int \! \df z\, \frac{1+z^2}{1-z}\, \ln z\,
    \Delta \mathcal{O}_{N-1}({\bf 0},{\bf p}_{\text{ap}},z,{\bf 0}) \biggr] \nn \\
    &\quad = (2^{N-1}-1)\,\frac{\alpha_s C_F}{\pi}\! 
   \Bigl[ \mathcal{L}_0^T({\bf p},\mu)\,\, \Delta {\cal F}_{N-1}(|{\bf p}|,|{\bf p}_{\text{ap}}|, \phi) + \delta^2({\bf p}) \, \Delta j_{N-1}   \Bigr]
\,,\nn\end{align}
again pulling out the tree-level factor $(2^{N-1}-1)$.
The coefficients $\Delta {\cal F}_{N-1}$ and $\Delta j_{N-1}$ are defined through the integrals 
\begin{align}
   \Delta {\cal F}_{N-1} &= \frac{1}{2^{N-1}-1} \int\!\df z\, \biggl(\frac{1+z^2}{1-z} + (z \leftrightarrow 1\!-\!z)\biggr)
    \Bigl\{[2^{N-1} \!-\! (2\!-\!z)^{N-1}] \nn \\
    &\quad \quad - [(1+z)^{N-1} \!-\! 1] \Bigr\}  \biggl[\Theta\Bigl(z_* \!<\!z\!<\!\frac12\Bigr) - \Theta\Bigl(\frac12 \!<\! z \!<\! z_* \Bigr)\biggr] \, , \label{eq:CT}\\
    \Delta j_{N-1} &= \frac{1}{2^{N-1}-1} \int\! \frac{\df \phi}{\pi} \int\!\df z\, \biggl(\frac{1+z^2}{1-z} + (z \leftrightarrow 1\!-\!z)\biggr) \, \ln z\, 
    \Bigl\{[2^{N-1} \!-\! (2\!-\!z)^{N-1}] \nn \\ \label{eq:Cdelta}
    &\quad \quad - [(1+z)^{N-1} \!-\! 1] \Bigr\}  \biggl[\Theta\Bigl(z_* \!<\!z\!<\!\frac12\Bigr) - \Theta\Bigl(\frac12 \!<\! z \!<\! z_* \Bigr)\biggr]\biggr|_{|{\bf p}|=0}  
\,,\end{align}
where the symmetrization $z \leftrightarrow 1-z$ has been moved to the splitting functions, for simplicity. 
The dependence of $\Delta {\cal F}_{N-1}$ on $|{\bf p}_{\text{ap}}|$ and $\phi$ is hiding in $z_*$. By contrast, $\Delta j_{N-1}$ does not depend on these because it is evaluated at $|{\bf p}| = 0$. In principle it has a dependence on $\phi$, because
\begin{align}
z_* \stackrel{|{\bf p}| \to 0}{=}
    \begin{cases}
        0 & 0 < \phi < \frac{\pi}{2}\,, \\
        1 & \frac{\pi}{2} < \phi < \pi\,,
    \end{cases}
\end{align}
as in this limit which particle is closest to the antipode only depends on $\phi$. However, at $|{\bf p}|= 0$ this angle becomes ill-defined, which is why we average over it in \eqref{eq:Cdelta}. We can also directly calculate $\Delta J_{N-1}$ analytically in the $\vert {\bf p}\vert \to 0$ limit, resulting in 
\begin{equation}
    \frac{\Delta J_{N-1}}{2^{N-1}-1} \xrightarrow{\vert {\bf p} \vert \to0} \Delta j_{N-1} \delta^2({\bf p}) \, ,
\end{equation}
implying that $\Delta {\mathcal F}_{N-1} \to 0$ in this limit and
\begin{equation}
   \Delta j_{-0.5} = 0.621 \,, \quad \Delta j_1 = 0\,, \quad \Delta j_2 = -\frac{1}{24} - \frac12 \ln 2\, , \quad \Delta j_3 = -\frac{9}{112} - \frac{27}{28} \ln 2\,.
\end{equation}
 We have verified these (mostly) analytical results for $\Delta j_{N-1}$  against our numerical calculation in \eqref{eq:Cdelta}, finding agreement. This limiting case thus provides us with a cross-check.

\begin{figure}
\begin{subfigure}{0.49\textwidth}
    \centering
    \includegraphics[width=\linewidth]{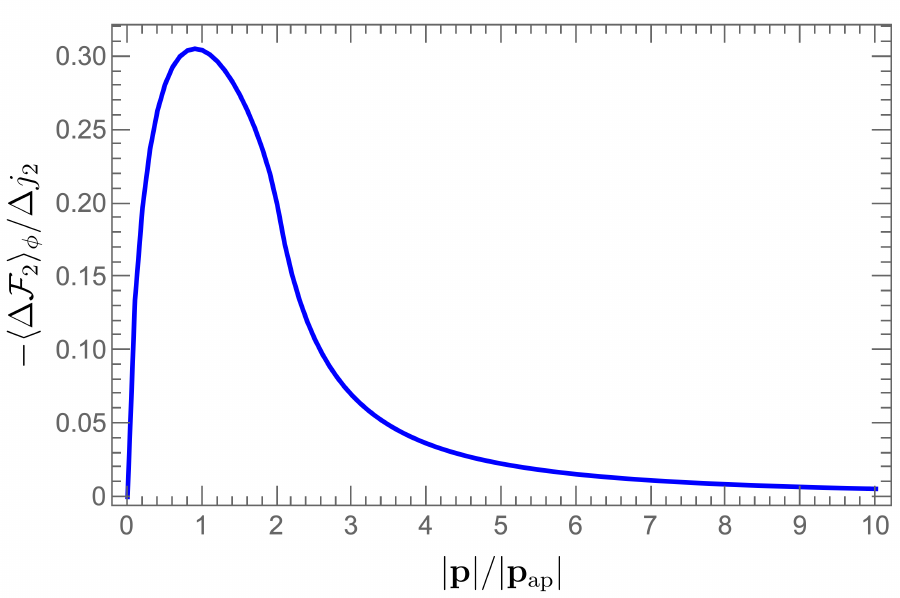}
    \caption{}
\end{subfigure}
\hfill
\begin{subfigure}{0.49\textwidth}
    \centering
    \includegraphics[width=\linewidth]{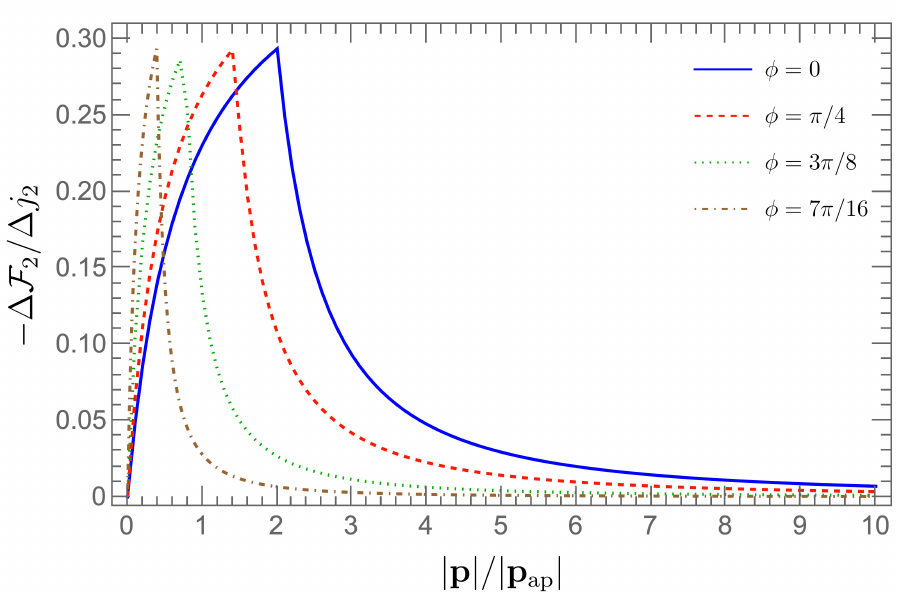}
    \caption{}
\end{subfigure}

\caption{
Coefficient $\Delta \mathcal{F}_{N-1}$ of the plus-distribution due to recoil in \eqref{eq:Delta_J} for $N=3$, (a) averaged over $\phi$ and (b) for specific $\phi$ values. The $N$-dependence is largely an overall factor.}
\label{fig:deltaJ}
\end{figure}

The remaining coefficient $\Delta {\cal F}_{N-1}$ of the plus-distribution in \eq{eq:Delta_J} is plotted in fig.~\ref{fig:deltaJ} as function of $|{\bf p}|/|{\bf p}_{\rm ap}|$ both for specific values of $\phi$ and averaged over $\phi$ (indicated by $\langle \cdots \rangle_\phi$).
In the $\phi$ dependent result there is a qualitative change due to the distinct cases in \eqref{eq:z_star}, which gets washed out when averaging over $\phi$.
We only show the result for $N=3$, as we have verified for a range of values of $N$ (including $N=0.5$) that
$\Delta {\cal F}_{N-1} /\Delta j_{N-1}$ is largely independent of $N$, up to corrections of a few percent. At the same time, we emphasize that $\Delta j_{N-1}/j_{N-1}$ changes with $N$.

\section{Non-perturbative effects in PENC distributions}
\label{sec:nonp}

Although PENCs obey simple scaling laws in perturbation theory, the non-perturbative power corrections can be sizable. For the two-point energy correlator, refs.~\cite{NASON1995291, Korchemsky:1999kt} demonstrated that leading corrections of order $\mathcal{O}(\Lambda_{\rm QCD}/Q)$ contribute across the entire range of angular scales, rather than being confined to the collinear and back-to-back region. In recent years, considerable effort has been devoted to understanding the size and structure of these leading $\mathcal{O}(\Lambda_{\rm QCD}/Q)$ corrections for EEC observables, employing both renormalon-based approaches as well as light-ray operator product expansion techniques~\cite{Schindler:2023cww, Lee:2024esz, Chen:2024nyc}. 

These approaches have been extended to (integer) PENCs in~\cite{Lee:2024esz, Chen:2024nyc}, while the expected $\Lambda_{\rm QCD}/Q$ structure of the analytically continued (non-integer) PENCs was first discussed in~\cite{Budhraja:2024tev}. In particular, ref.~\cite{Budhraja:2024tev} highlighted that for $N<1$ non-perturbative corrections become increasingly important and the leading corrections are no longer linear in $\Lambda_{\rm QCD}/Q$. In this study, we: (1)  use our new parametrization\footnote{The computation of (non-integer) PENCs in ref.~\cite{Budhraja:2024tev} relies on an approximate approach that uses sub-jets instead of final-state particles~\cite{Budhraja:2024xiq}.}, which allows for a realistic computation of (non-integer) PENCs with equal ease~\cite{Alipour-fard:2024szj}, (2) predict how the leading power-corrections depend on both the angular scale $x$ and the value of $N$ and (3) test this dependence across PENCs through the use of Monte Carlo simulations. 

Before discussing non-perturbative power corrections to PENCs, we would like to state an important caveat that caused us to limit our analysis to the collinear regime. In the traditional formulation, non-perturbative corrections to back-to-back configurations only impact the back-to-back region, i.e., $\chi \to \pi \sim \theta_{c_1,c_2}$, with $c_{1,2}$ representing collinear particles along the two different collinear directions in the back-to-back region. However, for our parametrization, depending on whether the special particle ($s$) is placed on the collinear particle $s \in {c_{1,2}}$ or a wide-angled soft particle $s \in {\rm soft_{NP}}$, two contributions arise. When the special particle is a collinear particle, the leading non-perturbative contributions with a single non-perturbative gluon still contribute to the $\chi \sim \pi$ angular regime. On the contrary, if the special particle is the non-perturbative gluon itself, these contributions now migrate to the fixed-order regime as $\theta_{s,{c_1/c2}} < \theta_{c_1,c_2} \sim \pi$. Therefore, sizable modifications due to non-perturbative effects can arise in the fixed-order region of the correlator, which we observe in \textsc{Pythia}. Consequently, we choose to focus only on the collinear regime for the rest of our discussion.

In the next section, we begin by discussing the leading order structure of the power corrections for any $N$-point energy correlator, and in sec.~\ref{sec:pythia} we test our analytic predictions against the hadronization model used in the {\textsc{Pythia}} Monte Carlo simulation~\cite{Bierlich:2022pfr}.

\subsection{Structure of leading non-perturbative power corrections}
The leading power correction to $N$-point energy correlators can be schematically written as~\cite{NASON1995291, Korchemsky:1999kt, Schindler:2023cww, Lee:2024esz, Chen:2024nyc}
\begin{equation}
    \frac{1}{\sigma}\frac{\df\sigma^{[N]}}{\df x}
    = \frac{1}{\sigma}\frac{\df\hat{\sigma}^{[N]}}{\df x}
    + \text{power corrections} \, ,
\end{equation}
where $\sigma^{[N]}$ denotes the full PENC distribution including non-perturbative effects, while $\hat{\sigma}^{[N]}$ represents the perturbative contribution. 
Though power corrections are relevant across the whole angular regime for energy correlators, their analytic structure turns out to be significantly simpler compared to standard di-jet observables.

Indeed, ref.~\cite{Schindler:2023cww} showed that employing the same leading power-correction parameter, $\Omega_1$, extracted from thrust data in the di-jet limit of $e^+e^-$ collisions~\cite{Abbate:2010xh} yields an excellent description of the EEC distribution in the existing $e^+\, e^-$ collider data~\cite{OPAL:1990reb, StDenis:1991svb, OPAL:1993pnw}. This agreement holds across most of the measured phase space without requiring additional fits, with deviations appearing only near the kinematic endpoints where higher-order effects become important.  
The leading non-perturbative correction  entering these analyses is obtained by performing a non-perturbative operator product expansion~\cite{Lee:2006nr,Mateu:2012nk}, resulting in a calculable coefficient and a non-perturbative matrix element $\Omega_1$ of soft Wilson lines, $Y_n , Y_{\bar{n}}$
\begin{equation}\label{eq:om1}
    \Omega_{1,k} = \frac{1}{N_k}\langle0| \bar{Y}_{\bar{n}}^{\dagger k} Y_{n}^{\dagger k} {\cal E}_T(0) Y_{n}^{k} \bar{Y}_{\bar{n}}^{k} |0 \rangle \, ,
\end{equation}
where $k=q,g$ denotes the color channel (corresponding to the fundamental or adjoint representation for the Wilson lines) and $N_q=N_c$, $N_g = N_c^2-1$~\cite{Stewart:2014nna, Ferdinand:2023vaf}. Here ${\cal E}_T$ is the boost-invariant transverse energy flow operator, which is related to the longitudinal energy flow operator as~\cite{Mateu:2012nk}
\begin{equation}\label{eq:etrans}
    \mathcal{E}_T = \frac{\mathcal{E}(\eta)}{\cosh \eta} \, .
\end{equation}
Employing similar arguments as in ref.~\cite{Lee:2024esz}, we now derive the leading power corrections for general $N$ within our parametrization. 

The leading non-perturbative correction arises when a single non-perturbative gluon is correlated with the collinear particles in the jet. For simplicity, consider a single collinear particle described by momentum fraction $z_c$ and a non-perturbative gluon carrying momentum fraction $z_{\rm NP}$, with $z_{\rm NP} \sim {\cal O}(\Lambda_{\rm QCD}/Q)$.\footnote{Technically, adding more collinear particles for the description of the leading non-perturbative effect, dictated by a single non-perturbative gluon, is straightforward as long as their relative angle is small compared to the angle with the non-perturbative gluon. However, to keep the expressions simple we focus on only a single collinear particle.} The contribution to the (differential) energy correlator from a collinear particle and a non-perturbative gluon then gives  
\begin{equation}\label{eq:dividing col soft}
    \bigg\{ z_c [(z_c+z_{\rm NP})^{N-1} - z_{c}^{N-1}] + z_{\rm NP}[(z_c + z_{\rm NP})^{N-1}- z_{\rm NP}^{N-1}] \bigg\} \, \delta(\chi-\theta_{{\rm NP},c})\, , 
\end{equation}
where $\theta_{{\rm NP},c}$ is the angle between the collinear particle and the non-perturbative gluon. In the first term, the special particle is the collinear particle while in the second the special particle is the non-perturbative gluon. 

Expanding the two terms, keeping only the leading terms in $z_{\rm NP}$ gives
\begin{align}\label{eq:np}
    \bigg\{ (N-1) z_c^{N-1} z_{\rm NP} + z_{\rm NP}[z_c^{N-1} - z_{\rm NP}^{N-1}] \bigg\} \, \delta(\chi-\theta_{{\rm NP},c}) \, .
\end{align}
The two $z_{\rm NP} \, z_c^{N-1}$ terms combine to produce the familiar leading non-perturbative corrections that grow linearly with $N$, similar to the result for (integer) $N \geq 2$ studied in the literature~\cite{Lee:2024esz, Chen:2024nyc}. The extra term proportional to $z_{\rm NP}^{N}$ is suppressed when $N > 1$ but provides the dominant contribution for $N<1$. As a result, the leading non-perturbative power corrections have a qualitatively different structure for $N<1$ than for $N>1$. Importantly, the result in \eq{eq:np} predicts that the non-perturbative corrections are \emph{negative} for $N<1$. In sec.~\ref{sec:pythia}, we will also find that this negative sign appears in the Monte Carlo simulations.

Rewriting $\eta$ in terms of the angle $\chi$ and then expressing that in terms of $x$ using \eq{eq:x} results in
\begin{equation}
  \cosh\eta = \frac{1}{2\sqrt{x(1-x)}} \sim \frac{1}{2 \sqrt{x}}  \, ,
\end{equation}
in the collinear limit. We use this to rewrite the energy flow in terms of the boost-invariant transverse energy flow operator, arriving at the following form for the leading non-perturbative corrections at leading order in $\alpha_s$\footnote{Interestingly, a similar dependence of the non-perturbative contributions on $\Lambda_{\rm QCD}/Q$ was observed for jet angularities in ref.~\cite{Larkoski:2013paa}. The connection of their findings to our results merits further investigation.}
\begin{align}\label{eq:sigma_NP_diff}
    \frac{1}{\sigma}\frac{\df\sigma_{\rm NP}^{[N]}}{\df x} &\equiv \frac{1}{\sigma}\frac{\df\sigma^{[N]}}{\df x} - \frac{1}{\sigma}\frac{\df\hat\sigma^{[N]}}{\df x} \, ,\nn \\
    &=  \frac{N\, \Omega_1}{2^N Q\,x^{3/2}}\,  - \frac{\tilde\Omega^{[N]}}{2^N Q^N\,x^{1+N/2}} \, .
\end{align}
Here $\df\sigma_{\rm NP}$ is the difference of the full PENC distribution and the perturbative computation and $x$, defined in \eq{eq:x}, is roughly $\chi^2/4$ in the collinear limit. We have additionally replaced the momentum fraction of the non-perturbative gluon with the parameter $\Omega_1$ (see \eq{eq:om1}) and the parameter $\tilde\Omega^{[N]}$ (discussed below). The factor of $1/2^N$ in the above equation arises due to normalization: we simulate $e^+\,e^- \to q {\bar{q}}$ events and hence the overall energy weights involved in the definition of PENC contain a factor of $Q/2$ for each particle. The factor of $1/2^N$ therefore ensures that the sum rule evaluates to 1.

For $N<1$, we find a new non-perturbative matrix element, which we denote as $\tilde\Omega^{[N]}$, that can be expressed as the matrix element of soft Wilson lines containing $N$ powers of energy flow operators as 
\begin{equation}\label{eq:omega_NP_smallN}
\tilde\Omega^{[N]} = \frac{1}{N_k} \langle0| \bar{Y}_{\bar{n}}^{\dagger k} Y_{n}^{\dagger k} {\cal E}_T^N(0) Y_{n}^{k} \bar{Y}_{\bar{n}}^{k} |0 \rangle\, , 
\end{equation}
We will discuss $\tilde\Omega^{[N]}$ in more detail in the next section. 
Due to the presence of $N$ powers of the non-perturbative gluon, for $N<1$, the dependence on the angular scale $x$ becomes non-trivial going as $x^{-1-N/2}$, where the $x^{-1}$ is just the classical scaling for differential energy correlators, see eq.~\eqref{eq:pert_coll_scaling}. Importantly, this implies that the non-perturbative power corrections approach the same classical scaling in $x$ as the perturbative term for $N \to 0$. 

\subsection{Assessing non-perturbative corrections using {\textsc{Pythia}}}
\label{sec:pythia}

We simulate electron-positron ($e^+e^-$) annihilation events using the \textsc{Pythia}~8.3 Monte Carlo event generator, considering the process $e^+e^- \rightarrow {\rm jets}$ at $Q= 1 \,{\rm TeV}$. The choice of $Q$ ensures a large collinear region of the distribution where our predictions for the non-perturbative contributions can be tested. Specifically, we want a sufficiently large region where the collinear factorization holds, while at the same time only including the leading non-perturbative corrections suffices. A total of $10^6$ events were used to compute the PENC distributions from the simulated data. In addition, to study the dependence of the power corrections on the center-of-mass energy, we also generated events at $Q = 500 \,{\rm GeV}$.

To determine the size of non-perturbative corrections in the simulations, we generate event samples both with and without hadronization effects enabled. The estimate of non-perturbative contributions from the Monte Carlo simulated data is then taken to be
\begin{equation}\label{eq:NP_MC}
    \frac{1}{\sigma}\frac{\df\sigma_{\rm NP, MC}^{[N]}}{\df x} = \frac{1}{\sigma}\frac{\df\sigma_{\rm hadron}^{[N]}}{\df x} - \frac{1}{\sigma}\frac{\df\sigma_{\rm parton}^{[N]}}{\df x} \, ,
\end{equation}
where the first term on the right is the hadron-level distribution of PENC and the second term is the parton-level distribution obtained by turning off hadronization in {\textsc{Pythia}}.

\begin{figure}
    \centering
    \includegraphics[scale=0.65]{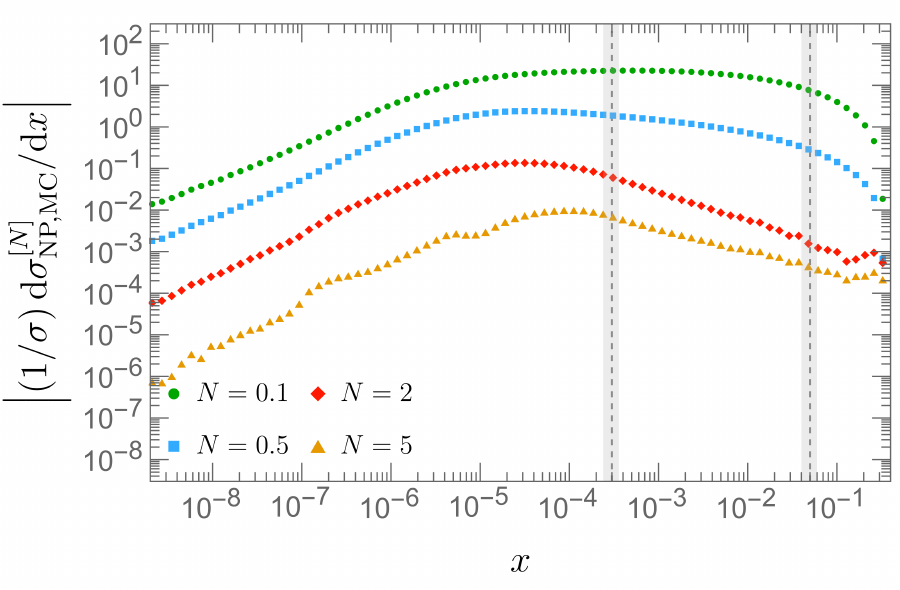}
    \caption{Log-log plot showing the non-perturbative contributions extracted from {\textsc{Pythia}} simulations for $N=0.1,0.5,2$ and $5$. The size of these corrections is significantly larger for $N<1$ than for $N>1$ values. Moreover, as can be seen when $N$ approaches $0$, the scaling region does not have a simple power law structure (it's not quite a straight line anymore). The vertical dashed lines represent the perturbative scaling region where leading non-perturbative corrections suffice and can be studied. The light gray band shows the variation of these boundaries by $\pm 20\%$, which will be used in fits.} 
    \label{fig:pc}
\end{figure}

\begin{figure}
    \centering
    \includegraphics[scale=0.65]{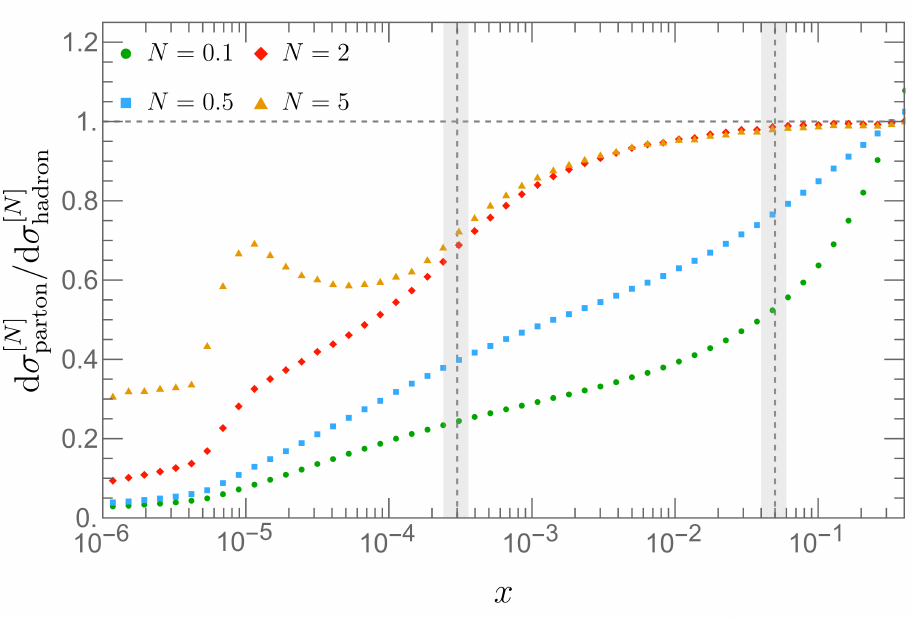}
    \caption{Ratio of parton level to hadron level distributions extracted from {\textsc{Pythia}} simulations for $N=0.1,0.5,2$ and $5$, illustrating the size of non-perturbative corrections to PENC distributions. While the non-perturbative corrections provide a sizable ($\sim {\cal O}(20\%)$) contribution for $N=2$ and $N=5$, they grow significantly in relevance for $N<1$. The figure is zoomed in to highlight the region where our soft gluon approximation is expected to hold.} 
    \label{fig:part_had}
\end{figure}

In fig.~\ref{fig:pc}, we plot the overall size of the non-perturbative contributions obtained from {\textsc{Pythia}} simulations, as defined by \eq{eq:NP_MC}, for $N=0.1, 0.5, 2$ and $5$. From the figure, we first observe that the overall size of power corrections is larger for smaller values of $N$, especially when $N<1$. We show the absolute value for this non-perturbative estimate as for $N<1$ the difference is negative, consistent with our analytic predictions (see \eq{eq:sigma_NP_diff}). The vertical dashed lines in fig.~\ref{fig:pc} indicate the fit region used for the estimation of the leading non-perturbative power corrections in our analysis. This region is selected such that the perturbative power law scaling holds in this regime and the non-perturbative corrections can be described sufficiently with the leading contribution. The shaded gray region in the figure shows the variation of this region by $\pm 20 \%$. To further illustrate the relevance of non-perturbative contributions to PENC distributions, we show the ratios of parton-level to hadron-level distributions extracted from \textsc{Pythia} simulations for $N=0.1, 0.5, 2,5$ in fig.~\ref{fig:part_had}. The figure demonstrates that non-perturbative corrections become parametrically large for $N<1$, consistent with our analytic predictions in eq.~\eqref{eq:sigma_NP_diff}. Below we will compare our analytic predictions for the whole range of $N$ values against \textsc{Pythia} simulations to provide an understanding of these corrections. 

We start by testing the predicted scaling behavior as function of $N$ for $N>1$ values. To this end, we take the ratios of non-perturbative corrections extracted from the simulations (see \eq{eq:NP_MC})  
to the corresponding quantity for the two-point energy correlator. From \eq{eq:sigma_NP_diff}, the leading non-perturbative power correction then scale as
\begin{equation}\label{Nscaling}
    \frac{\df\sigma_{\rm NP, MC}^{[N]}}{\df\sigma_{\rm NP, MC}^{[2]}} \approx \frac{\df\sigma_{\rm NP}^{[N]}}{\df\sigma_{\rm NP}^{[2]}} = \frac{N}{2^{N-1}} \, .
\end{equation}

\begin{figure}
    \centering
    \includegraphics[scale=0.65]{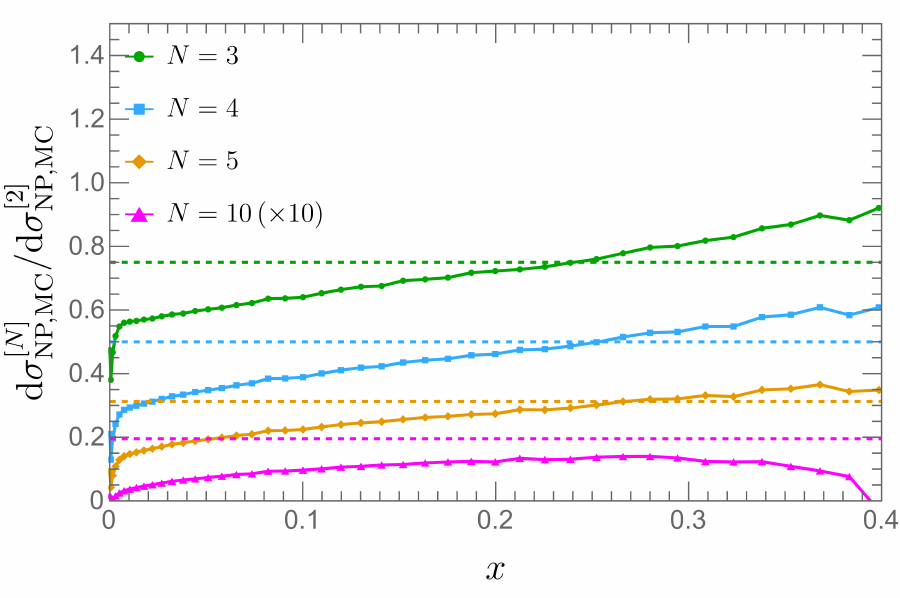}
    \caption{The difference between hadron level and parton level distributions, for $N = 3,\ 4,\ 5,\ 10$, normalized to the same quantity for $N = 2$. The horizontal dashed lines indicate leading order theoretical predictions. {\color{blue} The curve for $N=10$ has been multiplied by a constant factor of $10$ to enhance visibility.}}
    \label{N_dep_large}
\end{figure}
\begin{figure}[t]
    \centering
    \includegraphics[scale=0.7]{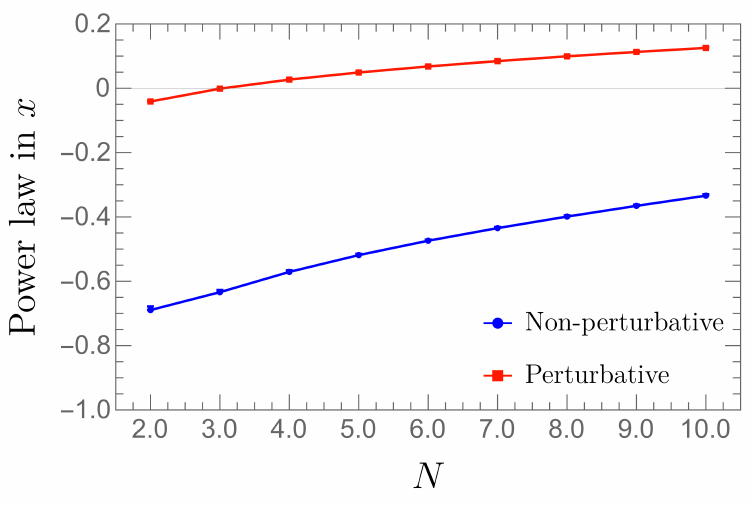}
    \caption{Extracted power law scaling from the log-log plots are shown as a function of $N$. Both perturbative (red) and non-perturbative (blue) power laws are shown. As $N$ increases, both move upward, reflecting the quantum scaling in $N$. The fit errors from varying the fit region are found to be tiny for $N>2$, causing the error bars to be barely visible.}
    \label{x_L_slopes}
\end{figure}

The resulting distributions are shown in fig.~\ref{N_dep_large} for $N = 3,\ 4,\ 5,$ and $10$ relative to $N=2$, plotted as a function of $x$. The horizontal dashed lines correspond to the leading order theoretical predictions. Due to the large suppression factor of $2^{N-1}$ from the overall normalization, we have multiplied the curve corresponding to $N=10$ by a constant factor of $10$ to make it more visible. From the figure we find that the leading power-corrections are close to the theoretically predicted values, but the simulated results from {\textsc{Pythia}} are not constant but depend on $x$. This shape of the power corrections emerges from the RGE of the  jet function in the collinear region, which induces an anomalous scaling governed by the \(N-1\) collinear partons~\cite{Lee:2024esz}. We will explore this further in fig.~\ref{x_L_slopes}.

\begin{figure}[t]
    \centering
    \includegraphics[scale=0.65]{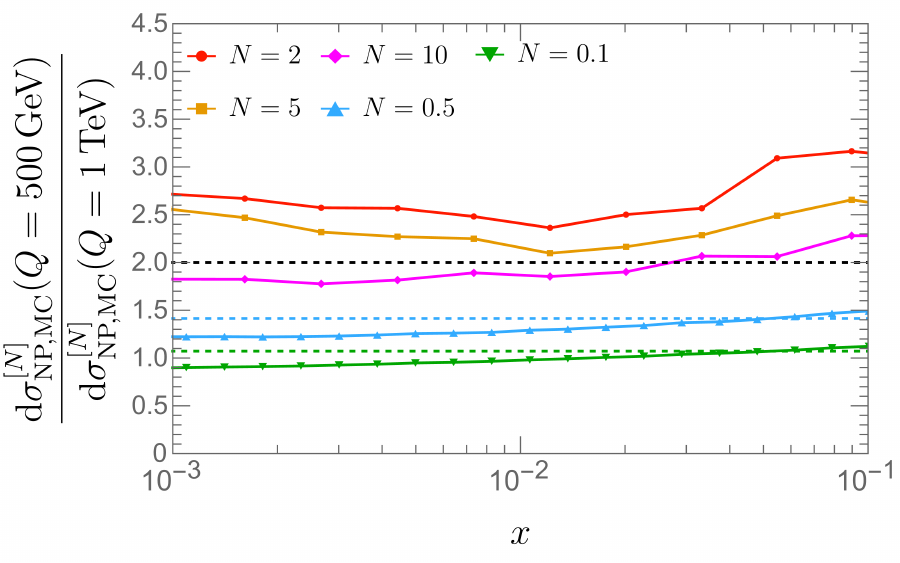}
    \caption{Ratio of the non-perturbative corrections at two different center-of-mass energies, $Q = 500\, {\rm GeV}$ and $Q = 1\, {\rm TeV}$ for various $N$-values ($N =  0.1,\ 0.5,\ 2,\ 5,\ 10$). The dashed lines represent the expected analytic prediction in \eqs{eq:Q_largeN}{eq:Q_smallN}. For $N>1$, deviations from the classical scaling are clearly visible, while for $N<1$ the simulated data is well-described by the predicted classical scaling.}
    \label{Q_dep}
\end{figure}

Next, we test the dependence on center-of-mass energy of these contributions by comparing the ratio of the non-perturbative corrections for a given $N$ at two different values of $Q$, namely $Q=500 \, {\rm GeV}$ and $Q=1 \, {\rm TeV}$. The expected scaling in $Q$ for $N>1$ is simply given as  
\begin{equation}\label{eq:Q_largeN}
     \frac{\df\sigma_{\rm NP, MC}^{[N]}(Q=500\, {\rm GeV})}{\df\sigma_{\rm NP,MC}^{[N]}(Q=1 \, {\rm TeV})} \approx \frac{\df\sigma_{\rm NP}^{[N]}(Q=500\, {\rm GeV})}{\df\sigma_{\rm NP}^{[N]}(Q=1 \, {\rm TeV})} = 2\, ,
\end{equation}
while for $N<1$, the leading power corrections exhibit a $Q$ dependence governed by
\begin{equation}\label{eq:Q_smallN}
    \frac{\df\sigma_{\rm NP, MC}^{[N]}(Q=500\, {\rm GeV})}{\df\sigma_{\rm NP,MC}^{[N]}(Q=1 \, {\rm TeV})} \approx \frac{\df\sigma_{\rm NP}^{[N]}(Q=500\, {\rm GeV})}{\df\sigma_{\rm NP}^{[N]}(Q=1 \, {\rm TeV})} = 2^{N}\, .
\end{equation}
The results are shown in fig.~\ref{Q_dep} for $N=0.1, 0.5, 2,5,10$. Although the predicted dependence on the center-of-mass energy is well described for $N<1$, the resulting distributions show clear deviations from the fixed-order (classical scaling) prediction for $N>1$ values. This could be attributed to the fact that for $N<1$, the overall size of non-perturbative effects is large and any violation of the classical scaling is a smaller effect than the differences between the classical scaling for different $N$. Meanwhile for $N>1$, the classical scaling is identical, so differences between different $N$ would be completely due to quantum corrections.

We will now test the dependence of the power corrections on the angular scale $x$ for $N>1$ correlators. From our leading order predictions, we expect this to scale as 
\begin{align}
    \log_{10} \biggl(\frac{1}{\sigma}\frac{\df\sigma_{\rm NP}^{[N]}}{\df \log_{10} x}\biggr) = -\tfrac12 \log_{10} x + \text{constant}\,
\end{align}
independent of $N$. To test this scaling, we use the region indicated in fig.~\ref{fig:pc} and fit the resulting distribution with a straight line to obtain the power law of the fitted distribution for different $N$ values. We show the extracted power law for both the parton-level distributions and non-perturbative contributions in fig.~\ref{x_L_slopes}. The fit errors are obtained by varying the fit region by $\pm 20\%$, they are shown in the figure but found to be small. From this figure, we observe that the extracted slope of the power corrections exhibits a dependence on $N$, contrary to the fixed-order predictions. Interestingly, we find that the $N$-dependence of this power law for the non-perturbative contributions has a similar shape as for the perturbative (parton-level) distributions, giving evidence of the modification of the power corrections through anomalous scaling and qualitatively following~\cite{Lee:2024esz}
\begin{equation}
    \gamma^{[N]}_{\rm NP} = -\frac{1}{2} + \hat\gamma^{[N-1]} \quad \text{for }N>1.
\end{equation}
Here $\gamma^{[N]}_{\rm NP}$ is the anomalous dimensions for the non-perturbative power corrections and $\hat\gamma^{[N-1]}$ is the anomalous scaling for the perturbative component for the $(N-1)$-point energy correlator. The extra $-1/2$ is the change in classical scaling for the leading non-perturbative corrections, see \eq{eq:sigma_NP_diff}.

\begin{figure}[t]
    \centering
    \includegraphics[scale=0.65]{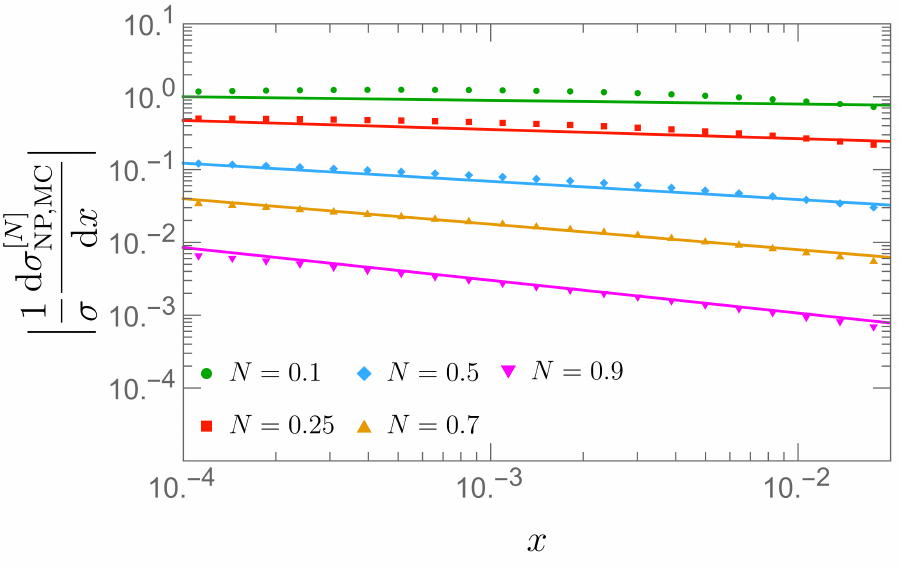}
    \caption{Log-log plot of the non-perturbative distribution, extracted via \eq{eq:NP_MC}, showing the scaling of power corrections with the angular scale $x$ for $N=0.1,0.25,0.5,0.7$ and $0.9$. The solid straight lines are the fits to a power law in $x$. The plot is zoomed to the fit region.} 
    \label{x_Ldep_smallN}
\end{figure}

\begin{figure}[t]
    \centering
    \includegraphics[scale=0.65]{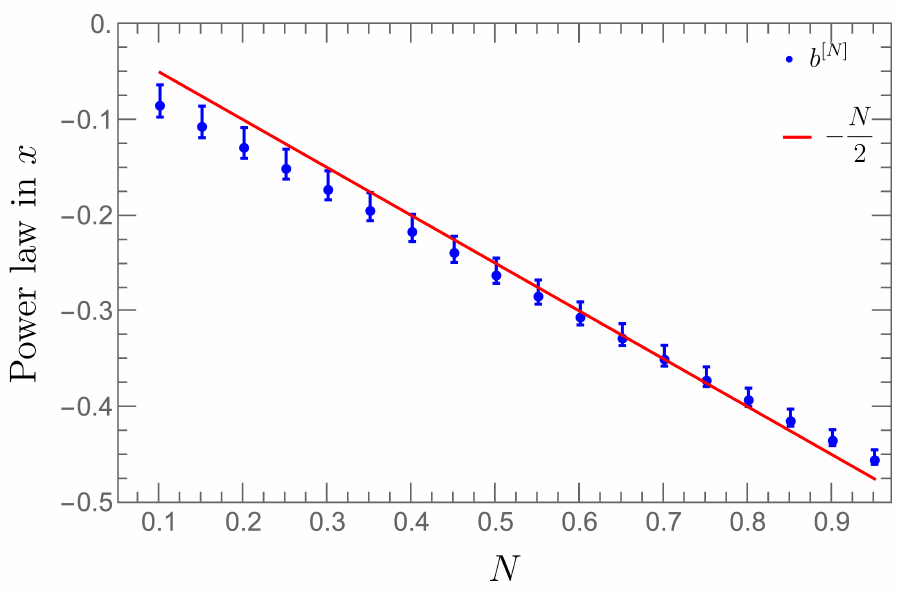}
    \caption{Extracted power law exponent for the dependence on the angular scale $x$ of the power corrections $N<1$ obtained from the fit in fig.~\ref{x_Ldep_smallN}. The solid red line shows the theoretical expectation from \eq{eq:x_smallN}.} 
    \label{x_L_exp_smallN}
\end{figure}

For $N < 1$, \eq{eq:sigma_NP_diff} dictates that the non-perturbative contributions exhibit a non-trivial $N$-dependence already at the leading order, given as
\begin{equation}\label{eq:x_smallN}
    \log_{10} \biggl(\frac{1}{\sigma}\frac{\df\sigma_{\rm NP}^{[N]}}{\df \log_{10} x}\biggr) = -\frac{N}{2} \log_{10} x + \text{constant}\,
\end{equation}
The fitted power law for different $N<1$ values is obtained by using the form 
\begin{equation}
\log_{10} \biggl(\frac{1}{\sigma}\frac{\df\sigma_{\rm fit}^{[N]\,;\, \rm NP,MC}}{\df \log_{10} x}\biggr) = a^{[N]} \,+\, b^{[N]} \log_{10} x \, 
\end{equation}
and is shown in fig.~\ref{x_Ldep_smallN} along with the distribution itself, zoomed in the fit region. From the figure we see that the fitted lines agree reasonably well in the scaling regime with the non-perturbative effects extracted from {\textsc{Pythia}} simulations. Next, in fig.~\ref{x_L_exp_smallN}, we compare the extracted power law $b^{[N]}$ from the fit as a function of $N$ with our theoretical prediction and find them to be in good agreement.


\begin{figure}[t]
    \centering
    \includegraphics[scale=0.65]{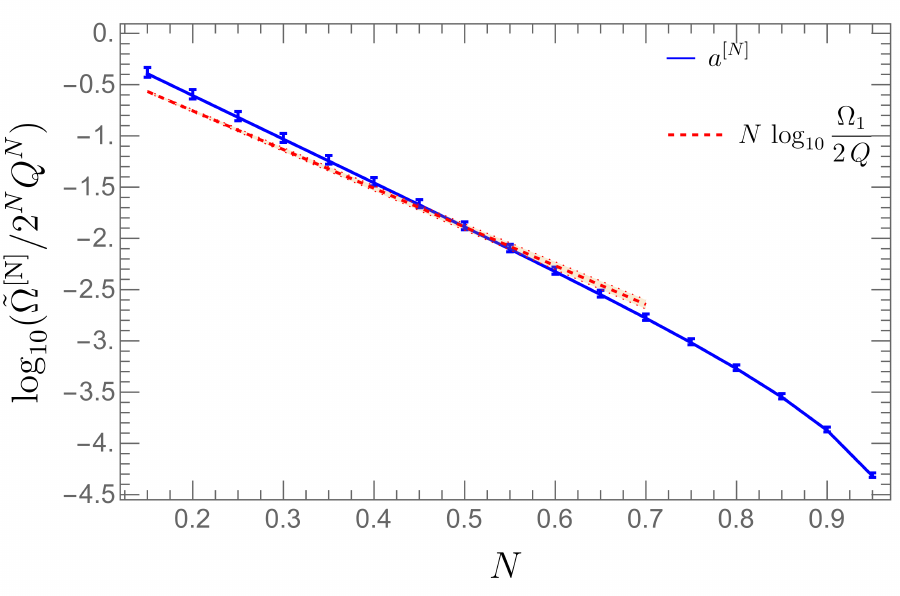}
    \caption{Extracted values of $\tilde\Omega^{[N]}$ from the fit compared against the theoretical expectation from a single soft non-perturbative emission.} 
    \label{omega_smallN}
\end{figure}
Finally, the constant term $a^{[N]}$ in the fit is related to the non-perturbative parameter $\tilde\Omega^{[N]}$ by
\begin{equation}
    a^{[N]} \sim \log_{10} \Big( \frac{\tilde\Omega^{[N]}}{2^N Q^N}\Big)\, .
\end{equation}
The result as a function of $N$ is shown in fig.~\ref{omega_smallN}. While in general $\langle z_{\rm NP}^{N}\rangle \neq \langle z_{\rm NP}\rangle^N$, if these expectation values are dominated by a single non-perturbative gluon, we expect $\tilde \Omega^{[N]}$ to equal $(\Omega_1)^N$. Comparing the extracted values from the fit, we find that $\tilde\Omega^{[N]}$ in this approximation is related to the $\Omega_1$ parameter as follows 
\begin{equation}
    \log_{10} \Big( \frac{\tilde\Omega^{[N]}}{2^N Q^N}\Big) 
    \approx N \log_{10} \Big( \frac{\Omega_1}{2\, Q}\Big) \, .
\end{equation}
We now use this approximation to fit $\Omega_1^{\textsc{Pythia}}$ in the range $0.15 \leq N \leq 0.65$ in fig.~\ref{omega_smallN}. With a goodness-of-fit measure $\chi^2/{\rm d.o.f} \approx 3$, the fit does not describe the data as well. However, the uncertainty on $\tilde \Omega^{[N]}$ values that we fit to are underestimated, as it only contains the effect of the variation of the fit region by $\pm 20\%$ (since the $\tilde \Omega^{[N]}$ values were obtained from a fit). The fit value of $\Omega_1^{\textsc{Pythia}} = 0.36\pm 0.06$ GeV is in the right ball park.
For example, the value $\Omega_1 = 0.305 \pm 0.084$ GeV was obtained from the thrust fit~\cite{Abbate:2010xh}, converted to the EEC in ref.~\cite{Schindler:2023cww}.
Therefore, we are not yet in the position to conclude that the single non-perturbative gluon approximation is violated for $N<1$, and thus the leading non-perturbative effect can be reasonably described by the same $\Omega_1$ as $N>1$. As $N$ approaches 1, the first term in \eq{eq:sigma_NP_diff} also contributes and we find that the result deviates from the scaling predicted by considering only the second term, as expected. It is also worth emphasizing that for $N<1$, these plots suggest that the RGE effects for the non-perturbative power corrections appear to be relatively small.  

\section{Conclusions}
\label{sec:summary}
In this work, we study projected $N$-point energy correlators (PENCs) exploiting the new parametrization introduced by some of us in ref.~\cite{Alipour-fard:2024szj}. Our parametrization organizes the correlator relative to a ``special particle'' thereby yielding a significantly simpler phase-space that allows computations to be naturally generalized to higher-point correlators in a unified framework. In contrast, for the traditional parametrization each PENC needs to be computed separately as the phase-space becomes increasingly complex for higher values of $N$. Utilizing this approach, we provide new results in two directions: 

First, we derive the factorization formula for general PENC distributions in the back-to-back limit. While the back-to-back limit of the two-point energy correlator has been extensively studied, higher-point correlators have so far not been studied in this regime. This computation is also motivated by experimental considerations as the back-to-back limit provides a complementary phase-space regime with different sources of systematics, that can potentially be used for future determinations of the strong coupling from energy correlators. A key result of our study is the computation of the one-loop jet function for PENCs. Our factorization formula is organized in a way such that the only non-trivial $N$-dependence resides in one of the jet functions (corresponding to the jet without the special particle in it). We also determine the impact of recoil induced by additional soft radiation that contributes in the back-to-back limit and complicates the determination of which particle is furthest from the special particle. Concretely, we show that the correction to the jet function due to recoil, denoted by $\Delta J_{N-1}$, is finite and can be computed numerically. Interestingly, we find that in $\Delta J_{N-1}$ the ratio of the coefficient of plus distribution $\Delta {\mathcal F}_{N-1}$ to the constant piece $\Delta j_{N-1}$ is largely independent of $N$, though the relative size of the recoil to the no-recoil result \emph{is} $N$-dependent. Our work provides all the necessary ingredients to carry out the NNLL resummation for higher-point energy correlators.     

Second, we study the analytic structure of leading non-perturbative power corrections to PENC distributions for arbitrary $N$ values, in the collinear limit. For $N > 1$, these power corrections scale linearly in N, consistent with earlier results in the literature~\cite{Lee:2024esz,Chen:2024nyc,Budhraja:2024tev}. However, we find a new non-perturbative contribution that dominates for $N < 1$ with a different dependence on the angular scale $x$, a non-trivial dependence on $N$, and involving a new non-perturbative matrix element $\tilde\Omega^{[N]}$.  Moreover, we find that the classical scaling is modified to $1/x^{1+N/2}$ for $N<1$, in contrast to $N>1$ where classical scaling remains $1/x^{3/2}$ for any $N$.
For $N \to 0$, this implies that the classical scaling in $x$ of the leading non-perturbative corrections is the same as the (fixed-order) perturbative power law scaling. Using a single soft-gluon approximation, we show that $\tilde\Omega^{[N]}$ can be related to the universal non-perturbative parameter $\Omega_1$ that enters in $e^{+}\, e^{-}$ event shape analyses (including PENCs with $N > 1$). Our analytic predictions are tested against the hadronization model in {\textsc{Pythia}}, finding qualitative agreement everywhere and quantitative agreement in many cases.

The results presented here pave the way for precision analyses with our new parametrization: In the back-to-back limit, the easier phase-space structure directly simplifies the factorization and makes the dependence on $N$ tractable. Non-perturbative corrections in the collinear region are not simplified by our new parametrization, but using our parametrization makes it testable on data, including for non-integer N.

\acknowledgments
We thank Hannah Bossi and Aditya Pathak for helpful discussions, and Johannes Michel and Jesse Thaler for their valuable feedback on the manuscript.

\bibliographystyle{elsarticle-num}
\bibliography{main}

@article{Ritzmann:2014mka,
    author = "Ritzmann, Mathias and Waalewijn, Wouter J.",
    title = "{Fragmentation in Jets at NNLO}",
    eprint = "1407.3272",
    archivePrefix = "arXiv",
    primaryClass = "hep-ph",
    reportNumber = "NIKHEF-2014-015",
    doi = "10.1103/PhysRevD.90.054029",
    journal = "Phys. Rev. D",
    volume = "90",
    number = "5",
    pages = "054029",
    year = "2014"
}

@article{Barata:2025zku,
    author = "Barata, Jo{\~a}o and Milhano, Jos{\'e} Guilherme and Sadofyev, Andrey V. and Silva, Jo{\~a}o M.",
    title = "{Early-Time Dynamics of Heavy-Ion Collisions through Energy Correlators: celestial blocks and the spacetime structure of out-of-equilibrium QCD matter}",
    eprint = "2512.17009",
    archivePrefix = "arXiv",
    primaryClass = "hep-ph",
    reportNumber = "CERN-TH-2025-257",
    month = "12",
    year = "2025"
}

@article{Kang:2024dja,
    author = "Kang, Zhong-Bo and Penttala, Jani and Zhang, Congyue",
    title = "{Determination of the strong coupling constant and the Collins-Soper kernel from the energy-energy correlator in $e^+e^-$ collisions}",
    eprint = "2410.21435",
    archivePrefix = "arXiv",
    primaryClass = "hep-ph",
    month = "10",
    year = "2024"
}

@article{Holguin:2026vld,
    author = "Holguin, Jack and Moult, Ian and Pathak, Aditya and Procura, Massimiliano and Sule, Siddharth",
    title = "{High precision heavy-boson-jet substructure with energy correlators}",
    eprint = "2601.20923",
    archivePrefix = "arXiv",
    primaryClass = "hep-ph",
    reportNumber = "DESY-26-012, UWThPh 2026-2",
    month = "1",
    year = "2026"
}

@article{Aglietti:2024xwv,
    author = "Aglietti, Ugo Giuseppe and Ferrera, Giancarlo",
    title = "{Energy-energy correlation in the back-to-back region at N3LL+NNLO in QCD}",
    eprint = "2403.04077",
    archivePrefix = "arXiv",
    primaryClass = "hep-ph",
    doi = "10.1103/PhysRevD.110.114004",
    journal = "Phys. Rev. D",
    volume = "110",
    number = "11",
    pages = "114004",
    year = "2024"
}

@article{Aglietti:2024zhg,
    author = "Aglietti, Ugo Giuseppe and Ferrera, Giancarlo",
    title = "{Heavy quark mass effects in the energy{\textendash}energy correlation in the back-to-back region}",
    eprint = "2412.02629",
    archivePrefix = "arXiv",
    primaryClass = "hep-ph",
    doi = "10.1140/epjc/s10052-025-13954-z",
    journal = "Eur. Phys. J. C",
    volume = "85",
    number = "3",
    pages = "272",
    year = "2025"
}

@article{Dokshitzer:1998kz,
    author = "Dokshitzer, Yuri L. and Lucenti, A. and Marchesini, G. and Salam, G. P.",
    title = "{On the QCD analysis of jet broadening}",
    eprint = "hep-ph/9801324",
    archivePrefix = "arXiv",
    reportNumber = "IFUM-602-FT",
    doi = "10.1088/1126-6708/1998/01/011",
    journal = "JHEP",
    volume = "01",
    pages = "011",
    year = "1998"
}

@article{Barata:2024wsu,
    author = "Barata, Jo{\~a}o and Kang, Zhong-Bo and Mayo L{\'o}pez, Xo{\'a}n and Penttala, Jani",
    title = "{Energy-Energy Correlator for Jet Production in pp and pA Collisions}",
    eprint = "2411.11782",
    archivePrefix = "arXiv",
    primaryClass = "hep-ph",
    reportNumber = "CERN-TH-2024-201",
    doi = "10.1103/96xh-bd1w",
    journal = "Phys. Rev. Lett.",
    volume = "134",
    number = "25",
    pages = "251903",
    year = "2025"
}

@article{Barata:2025uxp,
    author = "Barata, Jo{\~a}o and Brewer, Jasmine and Lee, Kyle and Silva, Jo{\~a}o M.",
    title = "{Heavy Quark Pair Energy Correlators: From Profiling Partonic Splittings to Probing Heavy-Flavor Fragmentation}",
    eprint = "2508.19404",
    archivePrefix = "arXiv",
    primaryClass = "hep-ph",
    reportNumber = "CERN-TH-2025-172, MIT-CTP 5888",
    month = "8",
    year = "2025"
}

@article{Duhr:2022yyp,
    author = "Duhr, Claude and Mistlberger, Bernhard and Vita, Gherardo",
    title = "{Four-Loop Rapidity Anomalous Dimension and Event Shapes to Fourth Logarithmic Order}",
    eprint = "2205.02242",
    archivePrefix = "arXiv",
    primaryClass = "hep-ph",
    reportNumber = "BONN-TH-2022-11, SLAC-PUB-17675",
    doi = "10.1103/PhysRevLett.129.162001",
    journal = "Phys. Rev. Lett.",
    volume = "129",
    number = "16",
    pages = "162001",
    year = "2022"
}

@article{Gao:2026xuq,
    author = "Gao, Anjie and Lee, Kyle and Zhang, Xiaoyuan",
    title = "{Precision Jet Substructure of Boosted Boson Decays with Energy Correlators}",
    eprint = "2601.20933",
    archivePrefix = "arXiv",
    primaryClass = "hep-ph",
    reportNumber = "MIT-CTP 5996, SLAC-PUB-260120",
    month = "1",
    year = "2026"
}

@article{Moult:2025nhu,
    author = "Moult, Ian and Zhu, Hua Xing",
    title = "{Energy Correlators: A Journey From Theory to Experiment}",
    eprint = "2506.09119",
    archivePrefix = "arXiv",
    primaryClass = "hep-ph",
    month = "6",
    year = "2025"
}

@article{Barata:2025fzd,
    author = "Barata, Jo{\~a}o and Moult, Ian and Sadofyev, Andrey V. and Silva, Jo{\~a}o M.",
    title = "{Dissecting Jet Modification in the QGP with Multi-Point Energy Correlators}",
    eprint = "2503.13603",
    archivePrefix = "arXiv",
    primaryClass = "hep-ph",
    reportNumber = "CERN-TH-2025-029",
    month = "3",
    year = "2025"
}

@article{Budhraja:2024tev,
    author = "Budhraja, Ankita and Chen, Hao and Waalewijn, Wouter J.",
    title = "{\ensuremath{\nu}-point energy correletors with FastEEC: Small-x physics from LHC jets}",
    eprint = "2409.12235",
    archivePrefix = "arXiv",
    primaryClass = "hep-ph",
    reportNumber = "MIT-CTP 5763",
    doi = "10.1016/j.physletb.2024.139239",
    journal = "Phys. Lett. B",
    volume = "861",
    pages = "139239",
    year = "2025"
}

@article{Singh:2024vwb,
    author = "Singh, Balbeer and Vaidya, Varun",
    title = "{Factorization for energy-energy correlator in heavy ion collision}",
    eprint = "2408.02753",
    archivePrefix = "arXiv",
    primaryClass = "hep-ph",
    doi = "10.1007/JHEP06(2025)071",
    journal = "JHEP",
    volume = "06",
    pages = "071",
    year = "2025"
}

@article{Holguin:2024tkz,
    author = {Holguin, Jack and Moult, Ian and Pathak, Aditya and Procura, Massimiliano and Sch\"ofbeck, Robert and Schwarz, Dennis},
    title = "{Top quark mass extractions from energy correlators: a feasibility study}",
    eprint = "2407.12900",
    archivePrefix = "arXiv",
    primaryClass = "hep-ph",
    reportNumber = "DESY-24-107;UWThPh 2024-14, DESY-24-107, UWThPh 2024-14",
    doi = "10.1007/JHEP04(2025)072",
    journal = "JHEP",
    volume = "04",
    pages = "072",
    year = "2025"
}

@article{Budhraja:2024xiq,
    author = "Budhraja, Ankita and Waalewijn, Wouter J.",
    title = "{FastEEC: Fast evaluation of N-point energy correlators}",
    eprint = "2406.08577",
    archivePrefix = "arXiv",
    primaryClass = "hep-ph",
    doi = "10.1016/j.physletb.2025.139276",
    journal = "Phys. Lett. B",
    volume = "861",
    pages = "139276",
    year = "2025"
}

@article{Lee:2024esz,
    author = "Lee, Kyle and Pathak, Aditya and Stewart, Iain W. and Sun, Zhiquan",
    title = "{Nonperturbative Effects in Energy Correlators: From Characterizing Confinement Transition to Improving \ensuremath{\alpha}s Extraction}",
    eprint = "2405.19396",
    archivePrefix = "arXiv",
    primaryClass = "hep-ph",
    reportNumber = "MIT-CTP 5711, DESY-24-064",
    doi = "10.1103/PhysRevLett.133.231902",
    journal = "Phys. Rev. Lett.",
    volume = "133",
    number = "23",
    pages = "231902",
    year = "2024"
}

@article{Jaarsma:2023ell,
    author = "Jaarsma, Max and Li, Yibei and Moult, Ian and Waalewijn, Wouter J. and Zhu, Hua Xing",
    title = "{Energy correlators on tracks: resummation and non-perturbative effects}",
    eprint = "2307.15739",
    archivePrefix = "arXiv",
    primaryClass = "hep-ph",
    doi = "10.1007/JHEP12(2023)087",
    journal = "JHEP",
    volume = "12",
    pages = "087",
    year = "2023"
}

@article{Li:2016axz,
    author = "Li, Ye and Neill, Duff and Zhu, Hua Xing",
    title = "{An exponential regulator for rapidity divergences}",
    eprint = "1604.00392",
    archivePrefix = "arXiv",
    primaryClass = "hep-ph",
    reportNumber = "FERMILAB-PUB-16-090-PPD-T, MIT-CTP-4795",
    doi = "10.1016/j.nuclphysb.2020.115193",
    journal = "Nucl. Phys. B",
    volume = "960",
    pages = "115193",
    year = "2020"
}

@article{CMS:2024mlf,
    author = "Hayrapetyan, Aram and others",
    collaboration = "CMS",
    title = "{Measurement of Energy Correlators inside Jets and Determination of the Strong Coupling \ensuremath{\alpha}S(mZ)}",
    eprint = "2402.13864",
    archivePrefix = "arXiv",
    primaryClass = "hep-ex",
    reportNumber = "CMS-SMP-22-015, CERN-EP-2024-010",
    doi = "10.1103/PhysRevLett.133.071903",
    journal = "Phys. Rev. Lett.",
    volume = "133",
    number = "7",
    pages = "071903",
    year = "2024"
}

@article{Chiu:2011qc,
    author = "Chiu, Jui-yu and Jain, Ambar and Neill, Duff and Rothstein, Ira Z.",
    title = "{The Rapidity Renormalization Group}",
    eprint = "1104.0881",
    archivePrefix = "arXiv",
    primaryClass = "hep-ph",
    doi = "10.1103/PhysRevLett.108.151601",
    journal = "Phys. Rev. Lett.",
    volume = "108",
    pages = "151601",
    year = "2012"
}

@article{Chiu:2009yx,
    author = "Chiu, Jui-yu and Fuhrer, Andreas and Hoang, Andre H. and Kelley, Randall and Manohar, Aneesh V.",
    title = "{Soft-Collinear Factorization and Zero-Bin Subtractions}",
    eprint = "0901.1332",
    archivePrefix = "arXiv",
    primaryClass = "hep-ph",
    doi = "10.1103/PhysRevD.79.053007",
    journal = "Phys. Rev. D",
    volume = "79",
    pages = "053007",
    year = "2009"
}

@article{Holguin:2023bjf,
    author = {Holguin, Jack and Moult, Ian and Pathak, Aditya and Procura, Massimiliano and Sch\"ofbeck, Robert and Schwarz, Dennis},
    title = "{Using the $W$ as a Standard Candle to Reach the Top: Calibrating Energy Correlator Based Top Mass Measurements}",
    eprint = "2311.02157",
    archivePrefix = "arXiv",
    primaryClass = "hep-ph",
    reportNumber = "UWThPh 2023-26; DESY-23-176",
    month = "11",
    year = "2023"
}

@article{Bauer:2003di,
    author = "Bauer, Christian W. and Lee, Christopher and Manohar, Aneesh V. and Wise, Mark B.",
    title = "{Enhanced nonperturbative effects in Z decays to hadrons}",
    eprint = "hep-ph/0309278",
    archivePrefix = "arXiv",
    doi = "10.1103/PhysRevD.70.034014",
    journal = "Phys. Rev. D",
    volume = "70",
    pages = "034014",
    year = "2004"
}

@article{Manohar:2003vb,
    author = "Manohar, Aneesh V.",
    title = "{Deep inelastic scattering as x ---{\ensuremath{>}} 1 using soft collinear effective theory}",
    eprint = "hep-ph/0309176",
    archivePrefix = "arXiv",
    doi = "10.1103/PhysRevD.68.114019",
    journal = "Phys. Rev. D",
    volume = "68",
    pages = "114019",
    year = "2003"
}

@article{Yang:2023dwc,
    author = "Yang, Zhong and He, Yayun and Moult, Ian and Wang, Xin-Nian",
    title = "{Probing the Short-Distance Structure of the Quark-Gluon Plasma with Energy Correlators}",
    eprint = "2310.01500",
    archivePrefix = "arXiv",
    primaryClass = "hep-ph",
    doi = "10.1103/PhysRevLett.132.011901",
    journal = "Phys. Rev. Lett.",
    volume = "132",
    number = "1",
    pages = "011901",
    year = "2024"
}

@article{Andres:2023ymw,
    author = "Andres, Carlota and Dominguez, Fabio and Holguin, Jack and Marquet, Cyrille and Moult, Ian",
    title = "{Seeing beauty in the quark-gluon plasma with energy correlators}",
    eprint = "2307.15110",
    archivePrefix = "arXiv",
    primaryClass = "hep-ph",
    doi = "10.1103/PhysRevD.110.L031503",
    journal = "Phys. Rev. D",
    volume = "110",
    number = "3",
    pages = "L031503",
    year = "2024"
}

@article{Chen:2024nyc,
    author = "Chen, Hao and Monni, Pier Francesco and Xu, Zhen and Zhu, Hua Xing",
    title = "{Scaling Violation in Power Corrections to Energy Correlators from the Light-Ray Operator Product Expansion}",
    eprint = "2406.06668",
    archivePrefix = "arXiv",
    primaryClass = "hep-ph",
    reportNumber = "CERN-TH-2024-084",
    doi = "10.1103/PhysRevLett.133.231901",
    journal = "Phys. Rev. Lett.",
    volume = "133",
    number = "23",
    pages = "231901",
    year = "2024"
}

@article{Barata:2023zqg,
    author = "Barata, Jo\~ao and Milhano, Jos\'e Guilherme and Sadofyev, Andrey V.",
    title = "{Picturing QCD jets in anisotropic matter: from jet shapes to energy energy correlators}",
    eprint = "2308.01294",
    archivePrefix = "arXiv",
    primaryClass = "hep-ph",
    doi = "10.1140/epjc/s10052-024-12514-1",
    journal = "Eur. Phys. J. C",
    volume = "84",
    number = "2",
    pages = "174",
    year = "2024"
}

@article{Chen:2023zlx,
    author = "Chen, Wen and Gao, Jun and Li, Yibei and Xu, Zhen and Zhang, Xiaoyuan and Zhu, Hua Xing",
    title = "{NNLL resummation for projected three-point energy correlator}",
    eprint = "2307.07510",
    archivePrefix = "arXiv",
    primaryClass = "hep-ph",
    doi = "10.1007/JHEP05(2024)043",
    journal = "JHEP",
    volume = "05",
    pages = "043",
    year = "2024"
}

@article{Schindler:2023cww,
    author = "Schindler, Stella T. and Stewart, Iain W. and Sun, Zhiquan",
    title = "{Renormalons in the energy-energy correlator}",
    eprint = "2305.19311",
    archivePrefix = "arXiv",
    primaryClass = "hep-ph",
    reportNumber = "MIT-CTP 5499",
    doi = "10.1007/JHEP10(2023)187",
    journal = "JHEP",
    volume = "10",
    pages = "187",
    year = "2023",
    note = "[Erratum: JHEP 10, 175 (2024)]"
}

@article{Barata:2023bhh,
    author = "Barata, Jo\~ao and Caucal, Paul and Soto-Ontoso, Alba and Szafron, Robert",
    title = "{Advancing the understanding of energy-energy correlators in heavy-ion collisions}",
    eprint = "2312.12527",
    archivePrefix = "arXiv",
    primaryClass = "hep-ph",
    reportNumber = "CERN-TH-2023-243",
    doi = "10.1007/JHEP11(2024)060",
    journal = "JHEP",
    volume = "11",
    pages = "060",
    year = "2024"
}

@article{Devereaux:2023vjz,
	Archiveprefix = {arXiv},
	Author = {Devereaux, Kyle and Fan, Wenqing and Ke, Weiyao and Lee, Kyle and Moult, Ian},
	Eprint = {2303.08143},
	Month = {3},
	Primaryclass = {hep-ph},
	Reportnumber = {MIT-CTP 5513, LA-UR-23-22542},
	Title = {{Imaging Cold Nuclear Matter with Energy Correlators}},
	Year = {2023}}

@article{Andres:2022ovj,
    author = "Andres, Carlota and Dominguez, Fabio and Kunnawalkam Elayavalli, Raghav and Holguin, Jack and Marquet, Cyrille and Moult, Ian",
    title = "{Resolving the Scales of the Quark-Gluon Plasma with Energy Correlators}",
    eprint = "2209.11236",
    archivePrefix = "arXiv",
    primaryClass = "hep-ph",
    doi = "10.1103/PhysRevLett.130.262301",
    journal = "Phys. Rev. Lett.",
    volume = "130",
    number = "26",
    pages = "262301",
    year = "2023"
}

@article{Craft:2022kdo,
	Archiveprefix = {arXiv},
	Author = {Craft, Evan and Lee, Kyle and Me\c{c}aj, Bianka and Moult, Ian},
	Eprint = {2210.09311},
	Month = {10},
	Primaryclass = {hep-ph},
	Reportnumber = {MIT-CTP 5474},
	Title = {{Beautiful and Charming Energy Correlators}},
	Year = {2022}}

@article{Kardos:2018kth,
	Archiveprefix = {arXiv},
	Author = {Kardos, Adam and Somogyi, G{\'a}bor and Tr{\'o}cs{\'a}nyi, Zolt{\'a}n},
	Doi = {10.1016/j.physletb.2018.10.014},
	Eprint = {1807.11472},
	Journal = {Phys. Lett.},
	Pages = {313-318},
	Primaryclass = {hep-ph},
	Slaccitation = {%%CITATION = ARXIV:1807.11472;%%},
	Title = {{Soft-drop event shapes in electron--positron annihilation at next-to-next-to-leading order accuracy}},
	Volume = {B786},
	Year = {2018}}

@article{Bauer:2001yt,
	Archiveprefix = {arXiv},
	Author = {Bauer, Christian W. and Pirjol, Dan and Stewart, Iain W.},
	Doi = {10.1103/PhysRevD.65.054022},
	Eprint = {hep-ph/0109045},
	Journal = {Phys. Rev. D},
	Pages = {054022},
	Reportnumber = {UCSD-PTH-01-15},
	Title = {{Soft collinear factorization in effective field theory}},
	Volume = {65},
	Year = {2002}}

@article{Abbate:2010xh,
	Archiveprefix = {arXiv},
	Author = {Abbate, Riccardo and Fickinger, Michael and Hoang, Andre H. and Mateu, Vicent and Stewart, Iain W.},
	Doi = {10.1103/PhysRevD.83.074021},
	Eprint = {1006.3080},
	Journal = {Phys. Rev.},
	Pages = {074021},
	Slaccitation = {%\%CITATION = 1006.3080;\%\%},
	Title = {{Thrust at N${}^3$LL with Power Corrections and a Precision Global Fit for $\alpha_s(m_Z)$}},
	Volume = {D83},
	Year = {2011}}

@article{Komiske:2022enw,
    author = "Komiske, Patrick T. and Moult, Ian and Thaler, Jesse and Zhu, Hua Xing",
    title = "{Analyzing N-Point Energy Correlators inside Jets with CMS Open Data}",
    eprint = "2201.07800",
    archivePrefix = "arXiv",
    primaryClass = "hep-ph",
    reportNumber = "MIT-CTP 5389",
    doi = "10.1103/PhysRevLett.130.051901",
    journal = "Phys. Rev. Lett.",
    volume = "130",
    number = "5",
    pages = "051901",
    year = "2023"
}

@article{Holguin:2022epo,
    author = "Holguin, Jack and Moult, Ian and Pathak, Aditya and Procura, Massimiliano",
    title = "{New paradigm for precision top physics: Weighing the top with energy correlators}",
    eprint = "2201.08393",
    archivePrefix = "arXiv",
    primaryClass = "hep-ph",
    doi = "10.1103/PhysRevD.107.114002",
    journal = "Phys. Rev. D",
    volume = "107",
    number = "11",
    pages = "114002",
    year = "2023"
}

@article{Ebert:2020sfi,
	Archiveprefix = {arXiv},
	Author = {Ebert, Markus A. and Mistlberger, Bernhard and Vita, Gherardo},
	Doi = {10.1007/JHEP08(2021)022},
	Eprint = {2012.07859},
	Journal = {JHEP},
	Pages = {022},
	Primaryclass = {hep-ph},
	Reportnumber = {MIT-CTP/5263, SLAC-PUB-17579, MPP-2020-225},
	Title = {{The Energy-Energy Correlation in the back-to-back limit at N$^{3}$LO and N$^{3}$LL'}},
	Volume = {08},
	Year = {2021}}

@article{Mateu:2012nk,
	Archiveprefix = {arXiv},
	Author = {Mateu, Vicent and Stewart, Iain W. and Thaler, Jesse},
	Doi = {10.1103/PhysRevD.87.014025},
	Eprint = {1209.3781},
	Journal = {Phys. Rev. D},
	Number = {1},
	Pages = {014025},
	Primaryclass = {hep-ph},
	Reportnumber = {IFIC-12-66, LPN12-099, MIT-CTP-4394},
	Title = {{Power Corrections to Event Shapes with Mass-Dependent Operators}},
	Volume = {87},
	Year = {2013}}

@article{Bierlich:2022pfr,
    author = "Bierlich, Christian and others",
    title = "{A comprehensive guide to the physics and usage of PYTHIA 8.3}",
    eprint = "2203.11601",
    archivePrefix = "arXiv",
    primaryClass = "hep-ph",
    reportNumber = "LU-TP 22-16, MCNET-22-04, FERMILAB-PUB-22-227-SCD",
    doi = "10.21468/SciPostPhysCodeb.8",
    journal = "SciPost Phys. Codeb.",
    volume = "2022",
    pages = "8",
    year = "2022"
}

@article{Lee:2006nr,
    author = "Lee, Christopher and Sterman, George F.",
    title = "{Momentum Flow Correlations from Event Shapes: Factorized Soft Gluons and Soft-Collinear Effective Theory}",
    eprint = "hep-ph/0611061",
    archivePrefix = "arXiv",
    reportNumber = "INT-PUB-06-17, YITP-SB-06-47",
    doi = "10.1103/PhysRevD.75.014022",
    journal = "Phys. Rev. D",
    volume = "75",
    pages = "014022",
    year = "2007"
}

@article{Stewart:2014nna,
	Archiveprefix = {arXiv},
	Author = {Stewart, Iain W. and Tackmann, Frank J. and Waalewijn, Wouter J.},
	Doi = {10.1103/PhysRevLett.114.092001},
	Eprint = {1405.6722},
	Journal = {Phys. Rev. Lett.},
	Number = {9},
	Pages = {092001},
	Primaryclass = {hep-ph},
	Reportnumber = {MIT-CTP-4530, DESY-14-008, NIKHEF-2014-002},
	Title = {{Dissecting Soft Radiation with Factorization}},
	Volume = {114},
	Year = {2015}}

@article{Basham:1979gh,
	Author = {Basham, C. Louis and Brown, Lowell S. and Ellis, Stephen D. and Love, Sherwin T.},
	Doi = {10.1016/0370-2693(79)90601-4},
	Journal = {Phys. Lett. B},
	Pages = {297--299},
	Reportnumber = {RLO-1388-786},
	Title = {{Energy Correlations in Perturbative Quantum Chromodynamics: A Conjecture for All Orders}},
	Volume = {85},
	Year = {1979}}

@article{Basham:1977iq,
	Author = {Basham, C. Louis and Brown, Lowell S. and Ellis, S. D. and Love, S. T.},
	Doi = {10.1103/PhysRevD.17.2298},
	Journal = {Phys. Rev. D},
	Pages = {2298},
	Reportnumber = {RLO-1388-746},
	Title = {{Electron - Positron Annihilation Energy Pattern in Quantum Chromodynamics: Asymptotically Free Perturbation Theory}},
	Volume = {17},
	Year = {1978}}

@article{Li:2021zcf,
    author = "Li, Yibei and Moult, Ian and van Velzen, Solange Schrijnder and Waalewijn, Wouter J. and Zhu, Hua Xing",
    title = "{Extending Precision Perturbative QCD with Track Functions}",
    eprint = "2108.01674",
    archivePrefix = "arXiv",
    primaryClass = "hep-ph",
    doi = "10.1103/PhysRevLett.128.182001",
    journal = "Phys. Rev. Lett.",
    volume = "128",
    number = "18",
    pages = "182001",
    year = "2022"
}

@article{Moult:2018jzp,
	Archiveprefix = {arXiv},
	Author = {Moult, Ian and Zhu, Hua Xing},
	Doi = {10.1007/JHEP08(2018)160},
	Eprint = {1801.02627},
	Journal = {JHEP},
	Pages = {160},
	Primaryclass = {hep-ph},
	Slaccitation = {%%CITATION = ARXIV:1801.02627;%%},
	Title = {{Simplicity from Recoil: The Three-Loop Soft Function and Factorization for the Energy-Energy Correlation}},
	Volume = {08},
	Year = {2018}}

@article{Collins:1981uk,
	Author = {Collins, John C. and Soper, Davison E.},
	Doi = {10.1016/0550-3213(81)90339-4},
	Journal = {Nucl. Phys. B},
	Note = {[Erratum: Nucl.Phys.B 213, 545 (1983)]},
	Pages = {381},
	Reportnumber = {OITS-155},
	Title = {{Back-To-Back Jets in QCD}},
	Volume = {193},
	Year = {1981}}

@article{Bauer:2001ct,
	Author = {Bauer, Christian W. and Stewart, Iain W.},
	Doi = {10.1016/S0370-2693(01)00902-9},
	Eprint = {hep-ph/0107001},
	Journal = {Phys. Lett.},
	Pages = {134-142},
	Reportnumber = {UCSD-PTH-01-09},
	Slaccitation = {%%CITATION = HEP-PH/0107001;%%},
	Title = {{Invariant operators in collinear effective theory}},
	Volume = {B516},
	Year = {2001}}

@article{Bauer:2002nz,
	Archiveprefix = {arXiv},
	Author = {Bauer, Christian W. and Fleming, Sean and Pirjol, Dan and Rothstein, Ira Z. and Stewart, Iain W.},
	Doi = {10.1103/PhysRevD.66.014017},
	Eprint = {hep-ph/0202088},
	Journal = {Phys. Rev.},
	Pages = {014017},
	Primaryclass = {hep-ph},
	Reportnumber = {UCSD-PTH-02-03},
	Slaccitation = {%%CITATION = HEP-PH/0202088;%%},
	Title = {{Hard scattering factorization from effective field theory}},
	Volume = {D66},
	Year = {2002}}

@article{Bauer:2000yr,
	Author = {Bauer, Christian W. and Fleming, Sean and Pirjol, Dan and Stewart, Iain W.},
	Doi = {10.1103/PhysRevD.63.114020},
	Eprint = {hep-ph/0011336},
	Journal = {Phys. Rev. D},
	Pages = {114020},
	Reportnumber = {UCSD-PTH-00-28},
	Slaccitation = {%%CITATION = HEP-PH/0011336;%%},
	Title = {{An Effective field theory for collinear and soft gluons: Heavy to light decays}},
	Volume = {63},
	Year = {2001}}

@article{Ji:2004wu,
    author = "Ji, Xiang-dong and Ma, Jian-ping and Yuan, Feng",
    title = "{QCD factorization for semi-inclusive deep-inelastic scattering at low transverse momentum}",
    eprint = "hep-ph/0404183",
    archivePrefix = "arXiv",
    reportNumber = "DOE-ER-40762-308, UM-PP-04-037, UM-PP-{\#}04-037",
    doi = "10.1103/PhysRevD.71.034005",
    journal = "Phys. Rev. D",
    volume = "71",
    pages = "034005",
    year = "2005"
}

@article{Chen:2021gdk,
    author = "Chen, Hao and Moult, Ian and Zhu, Hua Xing",
    title = "{Spinning gluons from the QCD light-ray OPE}",
    eprint = "2104.00009",
    archivePrefix = "arXiv",
    primaryClass = "hep-ph",
    doi = "10.1007/JHEP08(2022)233",
    journal = "JHEP",
    volume = "08",
    pages = "233",
    year = "2022"
}

@article{Chang:2013iba,
	Archiveprefix = {arXiv},
	Author = {Chang, Hsi-Ming and Procura, Massimiliano and Thaler, Jesse and Waalewijn, Wouter J.},
	Doi = {10.1103/PhysRevD.88.034030},
	Eprint = {1306.6630},
	Journal = {Phys. Rev.},
	Pages = {034030},
	Primaryclass = {hep-ph},
	Reportnumber = {MIT--CTP-4476},
	Slaccitation = {%%CITATION = ARXIV:1306.6630;%%},
	Title = {{Calculating Track Thrust with Track Functions}},
	Volume = {D88},
	Year = {2013}}

@article{Chang:2013rca,
	Archiveprefix = {arXiv},
	Author = {Chang, Hsi-Ming and Procura, Massimiliano and Thaler, Jesse and Waalewijn, Wouter J.},
	Doi = {10.1103/PhysRevLett.111.102002},
	Eprint = {1303.6637},
	Journal = {Phys. Rev. Lett.},
	Pages = {102002},
	Primaryclass = {hep-ph},
	Reportnumber = {MIT-CTP-4449, MIT--CTP-4449},
	Slaccitation = {%%CITATION = ARXIV:1303.6637;%%},
	Title = {{Calculating Track-Based Observables for the LHC}},
	Volume = {111},
	Year = {2013}}

@article{Chen:2019bpb,
	Archiveprefix = {arXiv},
	Author = {Chen, Hao and Luo, Ming-Xing and Moult, Ian and Yang, Tong-Zhi and Zhang, Xiaoyuan and Zhu, Hua Xing},
	Doi = {10.1007/JHEP08(2020)028},
	Eprint = {1912.11050},
	Journal = {JHEP},
	Number = {08},
	Pages = {028},
	Primaryclass = {hep-ph},
	Title = {{Three point energy correlators in the collinear limit: symmetries, dualities and analytic results}},
	Volume = {08},
	Year = {2020}}

@article{Korchemsky:1999kt,
	Archiveprefix = {arXiv},
	Author = {Korchemsky, Gregory P. and Sterman, George F.},
	Doi = {10.1016/S0550-3213(99)00308-9},
	Eprint = {hep-ph/9902341},
	Journal = {Nucl. Phys. B},
	Pages = {335--351},
	Reportnumber = {ITP-SB-98-73, LPT-ORSAY-98-80},
	Title = {{Power corrections to event shapes and factorization}},
	Volume = {555},
	Year = {1999}}

@article{Hofman:2008ar,
	Archiveprefix = {arXiv},
	Author = {Hofman, Diego M. and Maldacena, Juan},
	Doi = {10.1088/1126-6708/2008/05/012},
	Eprint = {0803.1467},
	Journal = {JHEP},
	Pages = {012},
	Primaryclass = {hep-th},
	Title = {{Conformal collider physics: Energy and charge correlations}},
	Volume = {05},
	Year = {2008}}

@article{Belitsky:2013xxa,
	Archiveprefix = {arXiv},
	Author = {Belitsky, A.V. and Hohenegger, S. and Korchemsky, G.P. and Sokatchev, E. and Zhiboedov, A.},
	Doi = {10.1016/j.nuclphysb.2014.04.020},
	Eprint = {1309.0769},
	Journal = {Nucl. Phys. B},
	Pages = {305--343},
	Primaryclass = {hep-th},
	Reportnumber = {CERN-PH-TH-2013-211, IPHT-T13-210, LAPTH-047-13},
	Title = {{From correlation functions to event shapes}},
	Volume = {884},
	Year = {2014}}

@article{Basham:1978bw,
	Author = {Basham, C.Louis and Brown, Lowell S. and Ellis, Stephen D. and Love, Sherwin T.},
	Doi = {10.1103/PhysRevLett.41.1585},
	Journal = {Phys. Rev. Lett.},
	Pages = {1585},
	Reportnumber = {RLO-1388-759},
	Title = {{Energy Correlations in electron - Positron Annihilation: Testing QCD}},
	Volume = {41},
	Year = {1978}}

@article{Basham:1978zq,
	Author = {Basham, C.L. and Brown, L.S. and Ellis, S.D. and Love, S.T.},
	Doi = {10.1103/PhysRevD.19.2018},
	Journal = {Phys. Rev. D},
	Pages = {2018},
	Reportnumber = {RLO-1388-761},
	Title = {{Energy Correlations in electron-Positron Annihilation in Quantum Chromodynamics: Asymptotically Free Perturbation Theory}},
	Volume = {19},
	Year = {1979}}

@article{Dixon:2019uzg,
	Archiveprefix = {arXiv},
	Author = {Dixon, Lance J. and Moult, Ian and Zhu, Hua Xing},
	Doi = {10.1103/PhysRevD.100.014009},
	Eprint = {1905.01310},
	Journal = {Phys. Rev. D},
	Number = {1},
	Pages = {014009},
	Primaryclass = {hep-ph},
	Reportnumber = {SLAC-PUB-17427, SLAC--PUB--17427},
	Title = {{Collinear limit of the energy-energy correlator}},
	Volume = {100},
	Year = {2019}}

@article{Chen:2020vvp,
	Archiveprefix = {arXiv},
	Author = {Chen, Hao and Moult, Ian and Zhang, XiaoYuan and Zhu, Hua Xing},
	Doi = {10.1103/PhysRevD.102.054012},
	Eprint = {2004.11381},
	Journal = {Phys. Rev. D},
	Number = {5},
	Pages = {054012},
	Primaryclass = {hep-ph},
	Title = {{Rethinking jets with energy correlators: Tracks, resummation, and analytic continuation}},
	Volume = {102},
	Year = {2020}}

@article{Fu:2024pic,
    author = {Fu, Yu and M{\"u}ller, Berndt and Sirimanna, Chathuranga},
    title = "{Modification of the Jet Energy-Energy Correlator in Cold Nuclear Matter}",
    eprint = "2411.04866",
    archivePrefix = "arXiv",
    primaryClass = "nucl-th",
    doi = "10.1103/qxdv-cmrg",
    journal = "Phys. Rev. Lett.",
    volume = "135",
    number = "11",
    pages = "112302",
    year = "2025"
}

@misc{Bossi:2024Nikhef,
  author = {Bossi, Hannah},
  title  = {{Energy correlators as a probe of QCD from simple to complex systems}},
  note   = {{Nikhef Theory Seminar, Amsterdam, Netherlands, Nov.~19, 2024}},
  year   = {2024},
  url    = {https://indico.nikhef.nl/event/6063/attachments/10747/16463/HBossi_NikhefSeminar.pdf}
}

@article{Singh:2025scb,
    author = "Singh, Balbeer",
    title = "{Toward factorization of jet observable in dense media: An EFT approach}",
    eprint = "2505.18070",
    archivePrefix = "arXiv",
    primaryClass = "hep-ph",
    doi = "10.1142/S0217751X25300121",
    journal = "Int. J. Mod. Phys. A",
    volume = "40",
    number = "28",
    pages = "2530012",
    year = "2025"
}

@article{Larkoski:2013paa,
    author = "Larkoski, Andrew J. and Thaler, Jesse",
    title = "{Unsafe but Calculable: Ratios of Angularities in Perturbative QCD}",
    eprint = "1307.1699",
    archivePrefix = "arXiv",
    primaryClass = "hep-ph",
    reportNumber = "MIT-CTP-4472, MIT--CTP-4472",
    doi = "10.1007/JHEP09(2013)137",
    journal = "JHEP",
    volume = "09",
    pages = "137",
    year = "2013"
}

@misc{Pathak:2024,
  author = {Pathak, Aditya},
  title  = {{Precision Top Mass Using
Energy Correlators}},
  note   = {{QCD@LHC, Oct, 2024}},
  year   = {2024},
  url    = {https://indico.cern.ch/event/1360294/contributions/6144507/attachments/2941385/5167878/Aditya_Pathak_QCD@LHC.pdf}
}

@book{Collins:2011zzd,
	Author = {Collins, John},
	Isbn = {978-1-107-64525-7, 978-1-107-64525-7, 978-0-521-85533-4, 978-1-139-09782-6},
	Month = {11},
	Publisher = {Cambridge University Press},
	Title = {{Foundations of perturbative QCD}},
	Volume = {32},
	Year = {2013}}

@article{Echevarria:2016scs,
    author = "Echevarria, Miguel G. and Scimemi, Ignazio and Vladimirov, Alexey",
    title = "{Unpolarized Transverse Momentum Dependent Parton Distribution and Fragmentation Functions at next-to-next-to-leading order}",
    eprint = "1604.07869",
    archivePrefix = "arXiv",
    primaryClass = "hep-ph",
    doi = "10.1007/JHEP09(2016)004",
    journal = "JHEP",
    volume = "09",
    pages = "004",
    year = "2016"
}

@article{Luo:2019nig,
	Archiveprefix = {arXiv},
	Author = {Luo, Ming-Xing and Shtabovenko, Vladyslav and Yang, Tong-Zhi and Zhu, Hua Xing},
	Doi = {10.1007/JHEP06(2019)037},
	Eprint = {1903.07277},
	Journal = {JHEP},
	Pages = {037},
	Primaryclass = {hep-ph},
	Title = {{Analytic Next-To-Leading Order Calculation of Energy-Energy Correlation in Gluon-Initiated Higgs Decays}},
	Volume = {06},
	Year = {2019}}

@article{Henn:2019gkr,
	Archiveprefix = {arXiv},
	Author = {Henn, J.M. and Sokatchev, E. and Yan, K. and Zhiboedov, A.},
	Doi = {10.1103/PhysRevD.100.036010},
	Eprint = {1903.05314},
	Journal = {Phys. Rev. D},
	Number = {3},
	Pages = {036010},
	Primaryclass = {hep-th},
	Reportnumber = {CERN-TH-2019-026, LAPTH-014/19, MPP-2019-54},
	Title = {{Energy-energy correlation in $N$=4 super Yang-Mills theory at next-to-next-to-leading order}},
	Volume = {100},
	Year = {2019}}

@article{Bossi:2024qho,
    author = "Bossi, Hannah and Kudinoor, Arjun Srinivasan and Moult, Ian and Pablos, Daniel and Rai, Ananya and Rajagopal, Krishna",
    title = "{Imaging the wakes of jets with energy-energy-energy correlators}",
    eprint = "2407.13818",
    archivePrefix = "arXiv",
    primaryClass = "hep-ph",
    reportNumber = "MIT-CTP-5739",
    doi = "10.1007/JHEP12(2024)073",
    journal = "JHEP",
    volume = "12",
    pages = "073",
    year = "2024"
}

@article{CMS:2024ovv,
    author = "CMS Collaboration",
    collaboration = "CMS",
    title = "{Energy-energy correlators from PbPb and pp collisions at 5.02 TeV}",
    journal = "Preprint",
    pages = "CMS-PAS-HIN-23-004",
    year = "2024"
}

@article{Tulipant:2017ybb,
    author = "Tulip{\'a}nt, Zolt{\'a}n and Kardos, Adam and Somogyi, G{\'a}bor",
    title = "{Energy{\textendash}energy correlation in electron{\textendash}positron annihilation at NNLL + NNLO accuracy}",
    eprint = "1708.04093",
    archivePrefix = "arXiv",
    primaryClass = "hep-ph",
    doi = "10.1140/epjc/s10052-017-5320-9",
    journal = "Eur. Phys. J. C",
    volume = "77",
    number = "11",
    pages = "749",
    year = "2017"
}

@article{Ferdinand:2023vaf,
    author = "Ferdinand, Anna and Lee, Kyle and Pathak, Aditya",
    title = "{Field-theoretic analysis of hadronization using soft drop jet mass}",
    eprint = "2301.03605",
    archivePrefix = "arXiv",
    primaryClass = "hep-ph",
    reportNumber = "MIT-CTP 5475, DESY-22-159",
    doi = "10.1103/PhysRevD.108.L111501",
    journal = "Phys. Rev. D",
    volume = "108",
    number = "11",
    pages = "L111501",
    year = "2023"
}

@inproceedings{StDenis:1991svb,
    author = "St. Denis, Richard Dante",
    collaboration = "ALEPH",
    title = "{Measurement of $\alpha_s$ from the structure of particle clusters produced in hadronic Z decays}",
    booktitle = "{26th Rencontres de Moriond: High-energy Hadronic Interactions}",
    pages = "171--177",
    year = "1991"
}

@article{OPAL:1990reb,
    author = "Akrawy, M. Z. and others",
    collaboration = "OPAL",
    title = "{A Measurement of energy correlations and a determination of $\alpha_s (M^2_{Z^0})$ in $e^+ e^-$ annihilations at $\sqrt{s} = 91$~GeV}",
    reportNumber = "CERN-PPE-90-121",
    doi = "10.1016/0370-2693(90)91098-V",
    journal = "Phys. Lett. B",
    volume = "252",
    pages = "159--169",
    year = "1990"
}

@article{CMS:2025jam,
    collaboration = "CMS",
    title = "{Energy-energy correlation in Z boson tagged PbPb and pp collisions at $\sqrt{s_{NN}}$ = 5.02 TeV}",
    reportNumber = "CMS-PAS-HIN-24-019",
    year = "2025"
}

@article{Becher:2011pf,
    author = "Becher, Thomas and Bell, Guido and Neubert, Matthias",
    title = "{Factorization and Resummation for Jet Broadening}",
    eprint = "1104.4108",
    archivePrefix = "arXiv",
    primaryClass = "hep-ph",
    reportNumber = "MZ-TH-11-08",
    doi = "10.1016/j.physletb.2011.09.005",
    journal = "Phys. Lett. B",
    volume = "704",
    pages = "276--283",
    year = "2011"
}

@article{Budhraja:2019mcz,
    author = "Budhraja, Ankita and Jain, Ambar and Procura, Massimiliano",
    title = "{One-loop angularity distributions with recoil using Soft-Collinear Effective Theory}",
    eprint = "1903.11087",
    archivePrefix = "arXiv",
    primaryClass = "hep-ph",
    reportNumber = "UWThPh 2019-5",
    doi = "10.1007/JHEP08(2019)144",
    journal = "JHEP",
    volume = "08",
    pages = "144",
    year = "2019"
}

@article{Becher:2011dz,
    author = "Becher, Thomas and Bell, Guido",
    title = "{Analytic Regularization in Soft-Collinear Effective Theory}",
    eprint = "1112.3907",
    archivePrefix = "arXiv",
    primaryClass = "hep-ph",
    doi = "10.1016/j.physletb.2012.05.016",
    journal = "Phys. Lett. B",
    volume = "713",
    pages = "41--46",
    year = "2012"
}

@article{Bauer:2002aj,
    author = "Bauer, Christian W. and Pirjol, Dan and Stewart, Iain W.",
    title = "{Factorization and endpoint singularities in heavy to light decays}",
    eprint = "hep-ph/0211069",
    archivePrefix = "arXiv",
    reportNumber = "UCSD-PTH-02-27, INT-PUB-02-49",
    doi = "10.1103/PhysRevD.67.071502",
    journal = "Phys. Rev. D",
    volume = "67",
    pages = "071502",
    year = "2003"
}

@article{Dokshitzer:1995zt,
    author = "Dokshitzer, Yuri L. and Webber, B. R.",
    title = "{Calculation of power corrections to hadronic event shapes}",
    eprint = "hep-ph/9504219",
    archivePrefix = "arXiv",
    reportNumber = "CAVENDISH-HEP-95-2, LU-TP-95-8",
    doi = "10.1016/0370-2693(95)00548-Y",
    journal = "Phys. Lett. B",
    volume = "352",
    pages = "451--455",
    year = "1995"
}

@article{Dokshitzer:1997ew,
    author = "Dokshitzer, Yuri L. and Webber, B. R.",
    title = "{Power corrections to event shape distributions}",
    eprint = "hep-ph/9704298",
    archivePrefix = "arXiv",
    reportNumber = "CAVENDISH-HEP-97-2",
    doi = "10.1016/S0370-2693(97)00573-X",
    journal = "Phys. Lett. B",
    volume = "404",
    pages = "321--327",
    year = "1997"
}

@article{Almeida:2014uva,
    author = "Almeida, Leandro G. and Ellis, Stephen D. and Lee, Christopher and Sterman, George and Sung, Ilmo and Walsh, Jonathan R.",
    title = "{Comparing and counting logs in direct and effective methods of QCD resummation}",
    eprint = "1401.4460",
    archivePrefix = "arXiv",
    primaryClass = "hep-ph",
    reportNumber = "LA-UR-13-28363, YITP-SB-14-02",
    doi = "10.1007/JHEP04(2014)174",
    journal = "JHEP",
    volume = "04",
    pages = "174",
    year = "2014"
}

@article{Collins:1984kg,
    author = "Collins, John C. and Soper, Davison E. and Sterman, George F.",
    title = "{Transverse Momentum Distribution in Drell-Yan Pair and W and Z Boson Production}",
    reportNumber = "CERN-TH-3923",
    doi = "10.1016/0550-3213(85)90479-1",
    journal = "Nucl. Phys. B",
    volume = "250",
    pages = "199--224",
    year = "1985"
}

@article{Collins:1981va,
    author = "Collins, John C. and Soper, Davison E.",
    title = "{Back-To-Back Jets: Fourier Transform from B to K-Transverse}",
    reportNumber = "OITS-153",
    doi = "10.1016/0550-3213(82)90453-9",
    journal = "Nucl. Phys. B",
    volume = "197",
    pages = "446--476",
    year = "1982"
}

@article{Jaarsma:2025tck,
    author = "Jaarsma, Max and Li, Yibei and Moult, Ian and Waalewijn, Wouter J. and Zhu, Hua Xing",
    title = "{From DGLAP to Sudakov: Precision Predictions for Energy-Energy Correlators}",
    eprint = "2512.11950",
    archivePrefix = "arXiv",
    primaryClass = "hep-ph",
    reportNumber = "MITP-24-091",
    month = "12",
    year = "2025"
}

@article{Electron-PositronAlliance:2025fhk,
    author = "Bossi, Hannah and others",
    collaboration = "Electron-Positron Alliance",
    title = "{Energy Correlators from Partons to Hadrons: Unveiling the Dynamics of the Strong Interactions with Archival ALEPH Data}",
    eprint = "2511.00149",
    archivePrefix = "arXiv",
    primaryClass = "hep-ph",
    reportNumber = "MITP-25-057, MITHIG-MOD-24-001",
    month = "10",
    year = "2025"
}

@article{OPAL:1993pnw,
    author = "Acton, P. D. and others",
    collaboration = "OPAL",
    title = "{A Determination of $\alpha_s(M_{Z^0})$ at LEP using resummed QCD calculations}",
    reportNumber = "CERN-PPE-93-38",
    doi = "10.1007/BF01555834",
    journal = "Z. Phys. C",
    volume = "59",
    pages = "1--20",
    year = "1993"
}

@article{Alipour-fard:2024szj,
    author = "Alipour-fard, Samuel and Budhraja, Ankita and Thaler, Jesse and Waalewijn, Wouter J.",
    title = "{New Angles on Energy Correlators}",
    eprint = "2410.16368",
    archivePrefix = "arXiv",
    primaryClass = "hep-ph",
    reportNumber = "MIT-CTP/5794",
    doi = "10.1103/l6nj-2gsh",
    journal = "Phys. Rev. Lett.",
    volume = "134",
    number = "23",
    pages = "231902",
    year = "2025"
}

@article{NASON1995291,
title = {Infrared renormalons and power suppressed effects in {$e^+e^-$} jet events},
journal = {Nuclear Physics B},
volume = {454},
number = {1},
pages = {291-309},
year = {1995},
doi = {https://doi.org/10.1016/0550-3213(95)00461-Z},
author = {Paolo Nason and Michael H. Seymour}
}

@article{STAR:2025jut,
    author = "Aboona, B. E. and others",
    collaboration = "STAR",
    title = "{Measurement of Two-Point Energy Correlators within Jets in $p+p$ Collisions at $\sqrt{s}=200$~GeV}",
    eprint = "2502.15925",
    archivePrefix = "arXiv",
    primaryClass = "hep-ex",
    doi = "10.1103/wv2t-dkgn",
    journal = "Phys. Rev. Lett.",
    volume = "135",
    number = "11",
    pages = "111901",
    year = "2025"
}

@article{CMS:2025ydi,
    author = "Chekhovsky, Vladimir and others",
    collaboration = "CMS",
    title = "{Observation of nuclear modification of energy-energy correlators inside jets in heavy ion collisions}",
    eprint = "2503.19993",
    archivePrefix = "arXiv",
    primaryClass = "nucl-ex",
    reportNumber = "CMS-HIN-23-004, CERN-EP-2025-014",
    doi = "10.1016/j.physletb.2025.139556",
    journal = "Phys. Lett. B",
    volume = "866",
    pages = "139556",
    year = "2025"
}

@article{ALICE:2025igw,
    author = "Acharya, Shreyasi and others",
    collaboration = "ALICE",
    title = "{Energy-energy correlators in charm-tagged jets in proton-proton collisions at $\mathbf{\sqrt{s} = 13}$ TeV}",
    eprint = "2504.03431",
    archivePrefix = "arXiv",
    primaryClass = "hep-ex",
    reportNumber = "CERN-EP-2025-082",
    month = "4",
    year = "2025"
}

@article{ALICE:2024dfl,
    author = "Acharya, Shreyasi and others",
    collaboration = "ALICE",
    title = "{Exposing the parton-hadron transition within jets with energy-energy correlators in pp collisions at $\sqrt{\textit s}=5.02$ TeV}",
    eprint = "2409.12687",
    archivePrefix = "arXiv",
    primaryClass = "hep-ex",
    reportNumber = "CERN-EP-2024-245",
    month = "9",
    year = "2024"
}

@article{Bossi:2024qeu,
    author = "Bossi, Hannah and Baty, Austin and Chen, Yi and Chen, Yu-Chen (Janice) and Innocenti, Gian-Michele and Maggi, Marcello and McGinn, Chris and Lee, Yen-Jie",
    title = "{Measurement of the energy-energy correlator in the back-to-back limit using the archived ALEPH e+e- data at 91.2 GeV}",
    eprint = "2501.01968",
    archivePrefix = "arXiv",
    primaryClass = "hep-ex",
    doi = "10.22323/1.478.0228",
    journal = "PoS",
    volume = "LHCP2024",
    pages = "228",
    year = "2025"
}

@article{Electron-PositronAlliance:2025wzh,
    author = "Bossi, Hannah and Chen, Yu-Chen and Chen, Yi and Zhang, Jingyu and Innocenti, Gian Michele and Badea, Anthony and Baty, Austin and Maggi, Marcello and McGinn, Chris and Lee, Yen-Jie",
    collaboration = "Electron-Positron Alliance",
    title = "{Analysis note: measurement of energy-energy correlator in $e^{+}e^{-}$ collisions at $91$ GeV with archived ALEPH data}",
    eprint = "2505.11828",
    archivePrefix = "arXiv",
    primaryClass = "hep-ex",
    month = "5",
    year = "2025"
}

@article{Xiao:2024rol,
    author = "Xiao, Meng and Ye, Yulei and Zhu, Xinyu",
    title = "{Prospect of measuring the top quark mass through energy correlators}",
    eprint = "2405.20001",
    archivePrefix = "arXiv",
    primaryClass = "hep-ph",
    doi = "10.1007/JHEP10(2024)088",
    journal = "JHEP",
    volume = "10",
    pages = "088",
    year = "2024"
}

@article{Budhraja:2025ulx,
    author = "Budhraja, Ankita and Singh, Balbeer",
    title = "{Exploiting {\ensuremath{\nu}}-dependence of projected energy correlators in HICs}",
    eprint = "2503.20019",
    archivePrefix = "arXiv",
    primaryClass = "hep-ph",
    doi = "10.1016/j.physletb.2025.140079",
    journal = "Phys. Lett. B",
    volume = "872",
    pages = "140079",
    year = "2026"
}

@article{deFlorian:2004mp,
    author = "de Florian, Daniel and Grazzini, Massimiliano",
    title = "{The Back-to-back region in e+ e- energy-energy correlation}",
    eprint = "hep-ph/0407241",
    archivePrefix = "arXiv",
    reportNumber = "CERN-PH-TH-2004-137",
    doi = "10.1016/j.nuclphysb.2004.10.051",
    journal = "Nucl. Phys. B",
    volume = "704",
    pages = "387--403",
    year = "2005"
}

\end{document}